\documentclass[10pt,aps,prd,fleqn,superscriptaddress,notitlepage,nofootinbib,preprintnumbers,nobalancelastpage]{revtex4-1}
\pdfoutput=1
\usepackage{amsmath,amssymb,graphicx,xspace,hyperref}
\usepackage[all]{hypcap}
\newcommand{\eps}{\varepsilon}

\usepackage{cancel,xcolor}

\hypersetup{hidelinks,
  pdfauthor={Stefan Hoeche, Matt LeBlanc,Jennifer Roloff,Grant Whitman},
  pdftitle={Massive tree-level splitting functions beyond kinematical limits},
  pdfkeywords={QCD,perturbation theory}
}

\begin{document}
\preprint{FERMILAB-PUB-25-0815-T, MCNET-25-28}
\title{Massive tree-level splitting functions beyond kinematical limits}
\author{Stefan~H{\"o}che}
\affiliation{Fermi National Accelerator Laboratory, Batavia, IL, 60510, USA}
\author{Matt LeBlanc}
\affiliation{Department of Physics, Brown University, Providence, RI, 02912, USA}
\author{Jennifer Roloff}
\affiliation{Department of Physics, Brown University, Providence, RI, 02912, USA}
\author{Grant Whitman}
\affiliation{Department of Physics, Brown University, Providence, RI, 02912, USA}

\begin{abstract}
We present a compact form of the massive $1\to 3$ tree-level QCD splitting functions
and discuss a decomposition of the results in terms of lower-order expressions,
scalar dipole antenna functions and pure higher-order remainders. The two-gluon
radiator functions introduced in this context are novel and generalize expressions
obtained from the double-soft approximation. Our results are obtained without
reference to soft or quasi-collinear limits.
\end{abstract}

\maketitle

\section{Introduction}
\noindent Final states with large multiplicities of jets initiated by massive partons
are of particular interest to the future of the Large Hadron Collider (LHC) 
experimental programme~\cite{ATL-PHYS-PUB-2025-018}. Both the top quark and Higgs boson
preferentially decay via bottom quarks. Higgs decays to charm quark pairs have also
grown in their relevance at the high-luminosity LHC (HL-LHC), following the rapid
development of jet flavor tagging algorithms powered by sophisticated
machine learning techniques~\cite{pmlr-v162-qu22b,FTAG-2023-05}.
Early versions of these heavy quark taggers primarily relied on identifying the
decay products of heavy hadrons produced in the jets, while modern taggers exploit
differences between the fragmentation patterns of light and heavy
partons~\cite{ATL-PHYS-PUB-2025-029,CMS-DP-2024-024,CMS-DP-2022-050,FTAG-2023-05}.
These algorithms have advanced to the point that processes originally thought
impossible to study at the HL-LHC -- namely, the production of Higgs boson pairs
decaying via bottom quarks -- will now be observed with high significance
before the end of Run~5~\cite{ATL-PHYS-PUB-2025-006}. The development of improved
theory predictions that include mass effects is thus highly motivated. 
At the same time, QCD jet evolution at the LHC has been measured to astonishing
precision~\cite{STDM-2017-04,CMS-SMP-16-010,CMS-SMP-20-010,STDM-2018-57,CMS-SMP-22-007,
  STDM-2023-07,ALICE:2021aqk,STDM-2022-05,ALICE:2024dfl,CMS-SMP-22-015,STDM-2017-33,ALICE:2025igw}.
While the large dynamic range originating in the high center-of-mass energy of the collider
makes it possible to treat the effects of heavy quark masses as a somewhat secondary
problem today, the ever increasing experimental precision will soon lead to scenarios
where massless approximations can no longer be justified. 
Improved theory for heavy quarks should also be included consistently within
the parton shower Monte Carlo algorithms~\cite{Buckley:2011ms,Campbell:2022qmc}
used both to measure jet dynamics and to train the fully supervised algorithms
that empower physics analyses. This will be necessary in particular to avoid
potential biases and uncertainty due to the model-dependence of
classifiers~\cite{JETM-2023-06,Gambhir:2025xim,GarciaCaffaro:2025gkm}.
In this work, we compute some important components needed for precision theory in
heavy quark final states: The tree-level three-parton splitting functions and the scalar
dipole antenna functions including mass effects. While the splitting functions are
already known~\cite{Dhani:2023uxu,Craft:2023aew}, we present them in a significantly
more compact form than existing literature, which will alleviate their integration
into massive infrared subtraction schemes, higher logarithmic resummation and
parton shower simulations, and provide faster evaluation and improved numerical stability
in practical applications. The massive scalar dipole radiators are computed 
for the first time, as is the decomposition of the splitting functions
into scalar radiators and pure splitting remainders.

The factorization of QCD scattering amplitudes is at the heart of most modern techniques
to compute higher-order QCD corrections to scattering processes at colliders. At fixed
order in perturbation theory, it allows one to construct infrared subtraction algorithms
that enable the extraction of singularities from the real and virtual corrections in
order to make each component finite and amenable to Monte-Carlo integration over the
final-state phase space~\cite{Heinrich:2020ybq,Huss:2022ful}.
In resummed perturbation theory, it enables the derivation of anomalous dimensions
that govern the scaling of cross sections with changing resolution scale.
For particles with masses, the factorization typically relies on the quasi-collinear
limit, which requires a simultaneous scaling of the collinear transverse momentum
and the parton masses. This procedure retains the correct soft behavior
of collinear scattering amplitudes at leading power, but it is somewhat unphysical,
because the on-shell parton masses should remain, in fact, constant.
As a direct consequence, one observes a non-commutativity of limits~\cite{
  Gaggero:2022hmv,Ghira:2023bxr,Ghira:2024nkk} that can only be resolved by accounting
for the physical features of the scattering matrix elements lost in the limiting procedure.
Here we use a different concept, which avoids the introduction of kinematical limits
altogether. It is based on the observation that infrared singularities arise
as a consequence of degenerate asymptotic states~\cite{Coleman:1965xm,Sterman:1995fz},
such that QCD splitting functions can be computed as reduced matrix elements in a
physical gauge~\cite{Frenkel:1976bia,Dokshitzer:1978hw,Ellis:1978sf,Ellis:1978ty,Humpert:1980te}.
Their leading singularities are determined by the semi-classical approximation~\cite{
  Gell-Mann:1954wra,Brown:1968dzy}, which can be derived from splitting functions in a
scalar QCD, with the scalars playing the role of the semi-classical particles.
This eventually allows one to derive any splitting function in terms of scalar multipole
radiator functions, and pure splitting remainders with a lower degree of divergence
in the kinematical limits~\cite{Campbell:2025lrs}.

This manuscript is organized as follows: In Sec.~\ref{sec:method} we provide
a brief introduction to the method and compute the one-to-two massive splitting
functions as simple examples. Section~\ref{sec:three-parton_tree-level} presents
the new results for one-to-three splitting functions with massive partons, as well
as the two-emission dipole antenna radiator functions with massive partons.
The decomposition of the splitting functions is discussed in
Sec.~\ref{sec:decomposition}, and Sec.~\ref{sec:outlook} presents an outlook.

\section{Background, techniques and leading-order results}
\label{sec:method}
The motivation for our work stems from the simple yet important observation
that the soft and collinear limits of QCD scattering amplitudes do not commute
beyond leading power. To exemplify this, we consider the one-to-two parton splitting
for a massive quark in the quasi-collinear limit. The splitting function reads
\begin{equation*}
  \langle P_{q\to qg}(z)\rangle=C_F\left[\frac{2z}{1-z}-\frac{m_q^2}{p_q p_g}
    +(1-\varepsilon)(1-z)\right]\,.
\end{equation*}
where $p_q$ and $p_g$ are the quark and gluon momenta and $z$ is the light-cone
momentum fraction of the quark with respect to the parent quark. The soft component
of this splitting function is given by the $z\to 1$ limit. At leading power in the
soft scaling parameter it reads
\begin{equation*}
    \langle P_{q\to q}(z)\rangle \overset{z\to 1}{\longrightarrow}
    C_F\bigg[\frac{2}{1-z}-\frac{m_q^2}{p_q p_g}\bigg]\;.
\end{equation*}
If this limit was physically meaningful, we should be able to first apply it
to a complete matrix element including the quark and the gluon, and then take the
quasi-collinear limit on the result. The leading-power soft limit of a
QCD matrix element is given in terms of the square of the soft-gluon current,
${\bf J}^\mu$~\cite{Bassetto:1984ik}.
For a process involving a heavy quark and a gluon in the final state,
the quark-gluon quasi-collinear limit on the squared current leads to
\begin{equation*}
    {\bf J}^\mu {\bf J}_\mu\overset{q\parallel g}{\longrightarrow}\,\propto
    C_F\bigg[\frac{2z}{1-z}-\frac{m_q^2}{p_qp_g}\bigg]\;.
\end{equation*}
The difference to the leading-power soft limit of the quasi-collinear result is
obvious. This simple example shows that the two kinematical limits do not commute,
even at the lowest non-trivial order. The reason is that leading terms in one of
the limits are sub-leading in the other. As a direct consequence, the leading-power
expressions obscure the factorization of higher-order QCD splitting functions,
and an extension to higher powers becomes mandatory to establish a systematic
factorization of matrix elements at higher orders in the coupling. We will therefore
refrain from using any kinematical approximation at all, and base
our calculation on a different principle.

The key observation of our method is that the leading components of splitting
functions in a gauge theory with spin are given in terms of the semi-classical
expressions, which coincide with the results of the corresponding scalar
theory~\cite{Gell-Mann:1954wra,Brown:1968dzy}. This allows us to separate out
the spin-dependent components, which do not contribute to a dipole radiation
pattern. The physical picture is consistent with the well-known decomposition
of the current into a charge current and a polarization current~\cite{Gordon:1928aa}.
In Ref.~\cite{Campbell:2025lrs}, this idea was used to define a new technique to
assemble splitting functions from scalar dipole radiators, lower-order splitting
functions, and pure higher-order splitting remainders. Here we will extend
this method to massive quarks. Because no collinear or soft limits
are taken, there is no question as to how to define their extension to the
quasi-collinear region, where the on-shell particle masses would have to be
scaled proportional to the transverse momentum. This in turn resolves common
problems which arise in the definition of polarizations of the external
particles and makes the splitting functions unambiguous.

It is important to note that the splitting remainders we compute are finite
in all hard collinear regions for fixed parton masses due to the dead-cone
effect~\cite{Marchesini:1989yk,Dokshitzer:1991fd,Ghira:2023bxr}.
Therefore, an infrared subtraction scheme can be devised without explicit knowledge
of those quantities. However, for small parton masses, QCD scattering matrix elements 
containing such splittings develop logarithms of the hard scale over the parton masses,
which should be resummed to correctly account for heavy jet dynamics. Treating such
logarithms on the same footing with standard collinear poles also improves the behavior
of (semi-)numerical fixed-order calculations~\cite{Catani:2002hc}.

\subsection{Tree-level techniques}
\label{sec:techniques}
In this section we introduce the techniques needed to perform the calculation
of the tree-level splitting functions. We also derive the lowest-order results,
which will be used in Sec.~\ref{sec:three-parton_tree-level} to decompose the
splitting functions at higher order. Similar approaches, working directly at
the level of helicity amplitudes, were introduced in~\cite{Duhr:2008wc,Cohen:2024xuf}.
To achieve a meaningful factorization of the results,
we use physical polarization states~\cite{
  Frenkel:1976bia,Dokshitzer:1978hw,Amati:1978wx,Amati:1978by,Ellis:1978sf,
  Ellis:1978ty,Kalinowski:1980ju,Kalinowski:1980wea,Humpert:1980te,Catani:1999ss}.
The polarization sum for virtual gluons is defined in terms of a
light-like axial gauge, which is known to be ghost free~\cite{
  Arnowitt:1962cv,Konetschny:1975he,Konetschny:1976im,
  Frenkel:1976zk,Konetschny:1979mw}
\begin{equation}\label{eq:axial_gauge}
    d^{\mu\nu}(p,\bar{n})=\sum_{\lambda=\pm}
    \varepsilon^\mu_\lambda(p,\bar{n})\varepsilon^{*\,\nu}_\lambda(p,\bar{n})
    =-g^{\mu\nu}+\frac{p^\mu \bar{n}^\nu+p^\nu \bar{n}^\mu}{p\bar{n}}\;.
\end{equation}
Here $\bar{n}^\mu$ is an auxiliary light-like vector. In the case of
on-shell external partons, the polarization states are always physical.
However, the factorized expressions in Sec.~\ref{sec:decomposition}
will sometimes involve external partons that are off mass shell. In such cases,
we construct the corresponding on-shell momentum needed to define physical
wave functions with the help of the auxiliary vector, $\bar{n}^\mu$.
For a parton with momentum $p^\mu$ and on-shell mass $m$, this shifted
momentum reads
\begin{equation}\label{eq:momentum_shift_massive}
  \bar{p}^\mu=p^\mu-\frac{p^2-m^2}{2p\bar{n}}\,\bar{n}^\mu\;.
\end{equation}
Our calculations are based on the techniques outlined in~\cite{Catani:1999ss}.
In this method, the splitting amplitudes for quarks or gluons into $l$ external
partons with momenta $p_1\ldots p_l$ are given by
\begin{equation}\label{eq:massive_splittings_recursive}
  \begin{split}
    P_{q}^{ss'}(1,\ldots,l)=&\;\delta^{ss'}
    \left(\frac{s_{1\ldots l}-m_{1\ldots l}^2}{
    8\pi\alpha_s\mu^{2\eps}}\right)^{l-1}
    \frac{{\rm Tr}\big[\,\slash\!\!\!\bar{n}
    \Psi(\{p_1,\ldots,p_l\})
    \bar{\Psi}(\{p_1,\ldots,p_l\})\,\big]}{
    {\rm Tr}\big[\,\slash\!\!\!\bar{n}
    \Psi(\{\bar{p}_{1\ldots l}\})
    \bar{\Psi}(\{\bar{p}_{1\ldots l}\})\,\big]}\;,\\
    P_{g}^{\mu\nu}(1,\ldots,l)=&\;\frac{D-2}{2}
    \left(\frac{s_{1\ldots l}}{
    8\pi\alpha_s\mu^{2\eps}}\right)^{l-1}
    \frac{d^{\mu\rho}(p_{1\ldots l},\bar{n})
    J_\rho(\{p_1,\ldots,p_l\})
    J_\sigma^{\dagger}(\{p_1,\ldots,p_l\})
    d^{\sigma\nu}(p_{1\ldots l},\bar{n})}{
    d^{\kappa\lambda}(\bar{p}_{1\ldots l},\bar{n})
    J_\lambda(\{\bar{p}_{1\ldots l}\})
    J_\tau^{\dagger}(\{\bar{p}_{1\ldots l}\})
    d^{\tau}_{\;\;\kappa}(\bar{p}_{1\ldots l},\bar{n})}\;.\\
  \end{split}
\end{equation}
Here, $s_{1\ldots l}=(p_1+\ldots+p_l)^2$ is the virtuality of the $l$-parton
final state, and $m_{1\ldots l}$ is the on-shell mass of the incoming parton.
The indices $s$ and $s'$ refer to the quark spin, while the gluon polarization
indices are given by $\mu$ and $\nu$.
The tree-level currents $\Psi$ and $J$ can be determined using recursive
techniques such as the Berends-Giele method~\cite{Berends:1987me,Berends:1988yn,
  Berends:1990ax,Duhr:2006iq}. We compute them in the light-like axial gauge,
Eq.~\eqref{eq:axial_gauge}. Only the expression for the quark current differs
from the massless case discussed in~\cite{Campbell:2025lrs}, and the difference
amounts to a trivial change in the quark propagator. We find
\begin{equation}\label{eq:quark_current}
  \begin{split}
    &\Psi_i(p_\alpha)=
    \sum_{\substack{\{\beta,\gamma\}\in\\P(\alpha,2)}}
    g_sT^a_{ij}\,%M^\mu(p_\beta,p_\gamma)
    \frac{i\sigma^{\mu\nu}}{p_{\alpha}^2-m_{\alpha}^2}\,p_{\gamma,\nu}\,
    J_\mu^a(p_\gamma,n)\Psi_j(p_\beta)\;,\\
    &\;\quad+\sum_{\substack{\{\beta,\gamma\}\in\\P(\alpha,2)}}
    \bigg[\,g_sT^a_{ij}S^\mu(p_\beta,p_\gamma) J_\mu^a(p_\gamma,n)
    -\!\!\!\sum_{\substack{\{\delta,\epsilon\}\in\\ OP(\gamma,2)}}
    \frac{g_s^2}{p_{\alpha}^2}\,\big\{T^a,T^b\big\}_{ij}\,
    J^{\mu,a}(p_\delta,n)J_\mu^b(p_\epsilon,n)\,\bigg]\Psi_j(p_\beta)\;,
  \end{split}
\end{equation}
\begin{equation}\label{eq:gluon_current}
  \begin{split}
    &J_\mu^a(p_\alpha,n)=
    \sum_{\substack{\{\beta,\gamma\}\in\\ P(\alpha,2)}}
    \bigg[\,g_s\frac{F^a_{bc}}{2}\,D_\mu(p_\beta,p_\gamma)
    J^{\rho,b}(p_\beta,n)J_\rho^c(p_\gamma,n)
    +g_sT^a_{ij}\,\bar{\Psi}_i(p_\gamma)
    d^{\mu\nu}(p_\alpha,n)\,\frac{\gamma_\nu}{p_\alpha^2}
    \Psi_j(p_\beta)\,\bigg]\\
    &\;\quad-\sum_{\substack{\{\beta,\gamma\}\in\\ P(\alpha,2)}}
    \bigg[\,g_sF^c_{ab}\,S^\sigma(p_\beta,p_\gamma) J_\sigma^c(p_\gamma,n)
    -\!\!\!\sum_{\substack{\{\delta,\epsilon\}\in\\ OP(\gamma,2)}}
    \frac{g_s^2}{p_{\alpha}^2}\big\{F^c,F^d\big\}_{ab}\,J^{\sigma,c}(p_\delta,n)
    J_\sigma^d(p_\epsilon,n)\,\bigg]\,
    d_\mu^{\;\;\nu}(p_\alpha,n)J_\nu^b(p_\beta,n)\;.
  \end{split}
\end{equation}
The expressions for the scalar production and decay vertices are
given in complete analogy to the massless case 
\begin{equation}\label{eq:tree_level_building_blocks}
  \begin{split}
    S^\mu(p_i,p_j)=&\;\frac{(2p_i+p_j)^\mu}{p_{ij}^2-m_{ij}^2}\;,
    \qquad\qquad
    &D^\mu(p_i,p_j,\bar{n})=&\;\frac{d^\mu_{\;\nu}(p_{ij},\bar{n})}{p_{ij}^2}\,(p_i-p_j)^\nu\;.
  \end{split}
\end{equation}
To obtain compact expressions for the splitting functions, it will be convenient
to use a Sudakov decomposition of the momenta~\cite{Sudakov:1954sw}.
We will restrict this decomposition to its most generic form, in particular
we will not rely on an explicit definition of the transverse components.
For two momenta, $p_i^\mu$ and $p_j^\mu$, we have
\begin{equation}\label{eq:sud_simple}
  \begin{split}
    p_i^\mu=&\;z_i\, p_{ij}^\mu
    -\tilde{p}_{i,j}^\mu\;,
    \qquad\text{and}\qquad
    &p_j^\mu=&\;z_j\, p_{ij}^\mu
    -\tilde{p}_{j,i}^\mu\;,
  \end{split}
\end{equation}
where $p_{ij}^\mu=p_i^\mu+p_j^\mu$.
The individual Sudakov components are
\begin{equation}\label{eq:def_z_ptilde}
    z_i=\frac{p_i\bar{n}}{p_{ij}\bar{n}}\;,
%    \qquad
%    z_j=\frac{p_j\bar{n}}{p_{ij}\bar{n}}\;,
    \qquad\text{and}\qquad
   \tilde{p}_{i,j}^\mu=\frac{z_ip_j^\mu-z_jp_i^\mu}{z_i+z_j}\;.
\end{equation}
This decomposition generalizes naturally to the many-parton case.

\subsection{Quark initial state}
\label{sec:two-parton_tree-level_quark}
\begin{figure}
  \centerline{\includegraphics[scale=0.36]{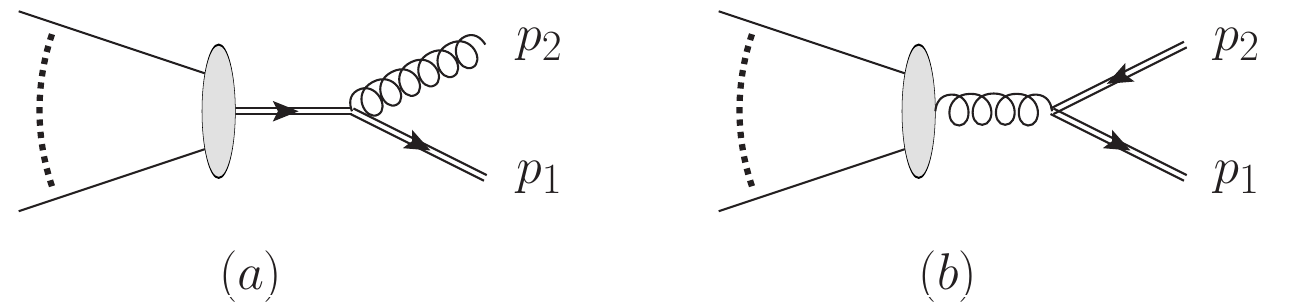}}
  \caption{Feynman diagrams contributing to the one-to-two parton splittings
    with massive quarks computed in Sec.~\ref{sec:two-parton_tree-level_quark}~(a)
    and Sec.~\ref{sec:two-parton_tree-level_gluon}~(b). Double lines indicate
    a quark of on-shell mass $m$.
  \label{fig:one-to-two_splittings}}
\end{figure}
We begin by deriving the splitting functions for one to two partons.
The corresponding diagrams are shown in Fig.~\ref{fig:one-to-two_splittings}.
Using Eq.~\eqref{eq:massive_splittings_recursive} and the recursive
definition of the quark current in Eq.~\eqref{eq:quark_current},
we find the following expression for the case of a quark to quark gluon
splitting
\begin{equation}\label{eq:coll_q_to_qg}
  \begin{split}
    P_{q\to q}(p_1,p_2)
    =&\;\frac{1}{8\pi\alpha_s\mu^{2\eps}}\,
    \frac{{\rm Tr}\big[\,\slash\!\!\!\bar{n}
    \Psi(\{p_1,p_2\})\bar{\Psi}(\{p_1,p_2\})\,\big]}{
    {\rm Tr}\big[\,\slash\!\!\!\bar{n}\Psi(\{\bar{p}_{12}\})
    \bar{\Psi}(\{\bar{p}_{12}\})\,\big]}
    =P_{\tilde{q}\to\tilde{q}}(p_1,p_2)+P_{q\to q}^{\rm(f)}(p_1,p_2)\;.
  \end{split}
\end{equation}
Its scalar and purely fermionic components are defined as
\begin{equation}\label{eq:coll_q_to_qg_components}
  \begin{split}
    P_{\tilde{q}\to\tilde{q}}(p_1,p_2)=&\;C_F\,\bigg[
    \frac{2z_1}{z_2}\bigg(1-\frac{p_1^2-m_1^2}{p_{12}^2-m_1^2}\frac{z_{12}}{z_1}
    -\frac{p_2^2}{p_{12}^2-m_1^2}\frac{z_{12}}{z_2}\bigg)
    -\frac{2m_1^2}{p_{12}^2-m_1^2}\,\bigg]\;,\\
    \langle P_{q\to q}^{\rm(f)}(p_1,p_2)\rangle
    =&\;C_F\,(1-\eps)\bigg(\frac{z_2}{z_{12}}
    -\frac{z_2}{z_1}\frac{p_1^2-m_1^2}{p_{12}^2-m_1^2}
      -\frac{p_2^2}{p_{12}^2-m_1^2}\bigg)\;,
  \end{split}
\end{equation}
These expressions are valid both for on-shell and for off-shell partons.
In order to achieve a physically meaningful definition of the external
wave functions in the off-shell region, we have used the shifted momentum
in Eq.~\eqref{eq:momentum_shift_massive} to compute the gluon polarization
tensor, $d^{\mu\nu}(p_2,\bar{n})\to d^{\mu\nu}(\bar{p}_2,\bar{n})=
  d^{\mu\nu}(p_2,\bar{n})-p_2^2\,\bar{n}^\mu\bar{n}^\nu/(p\bar{n})^2$.
In Sec.~\ref{sec:decomposition} we will also need the gluon-spin dependent
form of Eq.~\eqref{eq:coll_q_to_qg_components}. The complete quark to quark
splitting tensor is given by
\begin{equation}\label{eq:pqq_munu_sc}
  \begin{split}
    P_{q\to q}^{\mu\nu}(p_1,p_2)
    =&\;P_{\tilde{q}\to\tilde{q}}^{\mu\nu}(p_1,p_2)
    +P^{\rm(f)\mu\nu}_{q\to q}(p_1,p_2)\;.
  \end{split}
\end{equation}
The scalar and purely fermionic components are
\begin{equation}\label{eq:pqq_munu_sc_components}
  \begin{split}
    P_{\tilde{q}\to\tilde{q}}^{\mu\nu}(p_1,p_2)
    =&\;\frac{C_F}{2}\,(p_{12}^2-m^2)\,S^\mu(p_1,p_2)S^\nu(p_1,p_2)\;,\\
    P^{\rm(f)\mu\nu}_{q\to q}(p_1,p_2)
    =&\;-\frac{C_F}{2}\bigg[\,g^{\mu\nu}\bigg(\frac{z_2}{z_{12}}
    -\frac{z_2}{z_1}\frac{p_1^2-m^2}{p_{12}^2-m^2}
    -\frac{p_2^2}{p_{12}^2-m^2}\bigg)\\
    &\;\qquad\qquad+\frac{p_2^\mu p_2^\nu}{p_{12}^2-m^2}
    +\frac{\bar{p}_{12}^\mu \bar{p}_1^\nu-p_{12}^\mu p_1^\nu}{p_{12}^2-m^2}
    +\frac{\bar{p}_1^\mu\bar{p}_{12}^\nu-p_1^\mu p_{12}^\nu}{p_{12}^2-m^2}
    \,\bigg]\;,
  \end{split}
\end{equation}
Note that the scalar result is schematically identical
to the massless case~\cite{Campbell:2025lrs}.
It only differs in the denominator factors, which is
a trivial modification arising from the difference in
the retarded Green's function for massive particles.
The spin-dependent terms of the splitting function
exhibit minor modifications, which account for the change
in the off-shellness of the heavy quark.

\subsection{Gluon initial state}
\label{sec:two-parton_tree-level_gluon}
The second quark-mass dependent one-to-two reaction we need to compute is the
gluon to quark antiquark splitting shown in Fig.~\ref{fig:one-to-two_splittings}~(b).
Using Eq.~\eqref{eq:massive_splittings_recursive} and the recursive definition
of the current in Eq.~\eqref{eq:gluon_current}, we find the following expression
\begin{equation}\label{eq:coll_gqq_ggg_step1}
  \begin{split}
    P^{\mu\nu}_{g\to q}(p_1,p_2)
    =&\;\frac{T_R}{2p_{12}^2}\,d^\mu_{\;\rho}(p_{12},\bar{n})
    {\rm Tr}[\,(\slash\!\!\!\bar{p}_1\pm m)
    \gamma^\rho (\slash\!\!\!\bar{p}_2\mp m)\gamma^\sigma\,]
    d^\nu_{\;\sigma}(p_{12},\bar{n})\;.
  \end{split}
\end{equation}
A straightforward calculation, making use of the fact that
the final-state masses must be identical in QCD, gives
\begin{equation}\label{eq:coll_gqq}
  \begin{split}
    P^{\mu\nu}_{g\to q}(p_1,p_2)
    =&\;T_R\,\bigg[\,d^{\mu\nu}(\bar{p}_{12},\bar{n})
    \bigg(1-\frac{z_{12}}{p_{12}^2}
    \bigg(\frac{p_1^2-m^2}{z_1}+\frac{p_2^2-m^2}{z_2}\bigg)\bigg)\\
    &\;\qquad\quad+p_{12}^2\frac{\bar{n}^\mu\bar{n}^\nu}{(p_{12}\bar{n})^2}
    -p_{12}^2D^\mu(p_1,p_2,\bar{n})D^\nu(p_1,p_2,\bar{n})\,\bigg]\;,
  \end{split}
\end{equation}
In the on-shell case, and upon using a standard Sudakov
decomposition of the momenta~\cite{Sudakov:1954sw},
Eq.~\eqref{eq:coll_gqq} reduces to the familiar (massless)
DGLAP splitting kernel. It is interesting to observe
that the effects of parton masses in this particular case
are purely kinematical. Note that Eq.~\eqref{eq:coll_gqq}
is still independent of the kinematics parametrization.
Its spin averaged form reads
\begin{equation}\label{eq:coll_gqq_avg}
  \langle P_{g\to q}(p_1,p_2)\rangle
  =T_R\,\bigg[\,1-\frac{z_{12}}{p_{12}^2}
    \bigg(\frac{p_1^2-m^2}{z_1}+\frac{p_2^2-m^2}{z_2}\bigg)
    -\frac{2}{1-\eps}\bigg(\frac{z_1z_2}{z_{12}^2}
  -\frac{z_1p_1^2+z_2p_2^2}{z_{12}p_{12}^2}\bigg)\,\bigg]\;.
\end{equation}
We will use the above functions in Sec.~\ref{sec:decomposition}
to write the higher-order splitting functions in terms of
lower-order results.

\section{One-to-three splitting functions}
\label{sec:three-parton_tree-level}
In this section we provide the expressions for the one to three
parton splittings. Our calculations are based on the methods
introduced in~\cite{Catani:1999ss,Campbell:2025lrs}, in particular
the use of the light-like axial gauge, Eq.~\eqref{eq:axial_gauge}.
We employ Eqs.~\eqref{eq:massive_splittings_recursive} and
Eqs.~\eqref{eq:quark_current} and~\eqref{eq:gluon_current}
to determine the splitting functions in terms of Berends-Giele
currents. All results computed in this subsection are known,
however, we present them in a more compact form than the
existing literature. We have cross-checked the expressions
against Refs.~\cite{Dhani:2023uxu,Craft:2023aew} and verified
that they reduce to the massless case derived in
Refs.~\cite{Campbell:1997hg,Catani:1999ss} when the
parton masses are set to zero.

\subsection{All-quark splitting functions}
\label{sec:one_to_three_splittings_quark}
\begin{figure}
  \centerline{\includegraphics[scale=0.36]{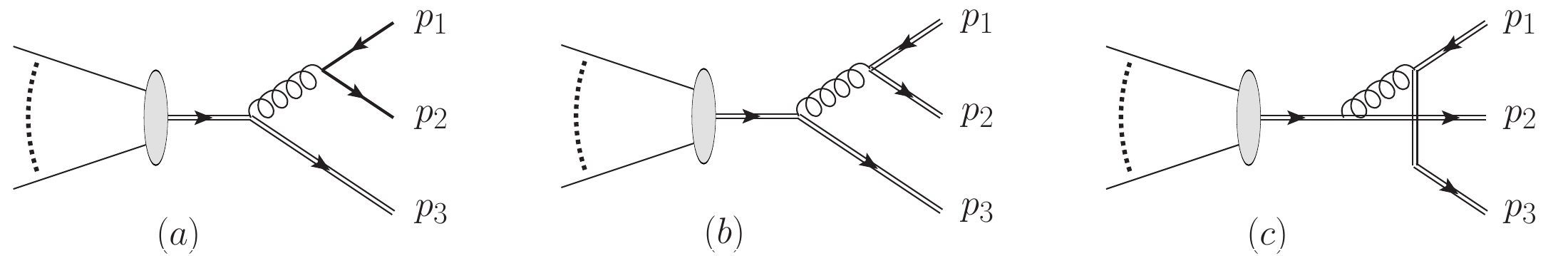}}
  \caption{Feynman diagrams contributing to the one-to-three all-quark splittings
    with massive quarks computed in Sec.~\ref{sec:one_to_three_splittings_quark}.
    Diagram~(a) shows a process with different on-shell masses for partons $1/2$
    and $3$, (indicated by the double line and the thick line in the figure),
    while diagrams~(b) and~(c) show a process with identical on-shell masses
    for all the quarks.
  \label{fig:one-to-three_splittings_quark}}
\end{figure}
The flavor-changing quark-to-quark splitting function can be obtained
from the product of the off-shell quark splitting function in
Eq.~\eqref{eq:pqq_munu_sc} and the spin-dependent gluon-to-quark splitting
function, Eq.~\eqref{eq:coll_gqq}. The corresponding Feynman diagram
is shown in Fig.~\ref{fig:one-to-three_splittings_quark}~(a). Note that
we do not make any assumptions about the on-shell quark masses, in particular
$m_3$ can differ from $m_1$. However, in QCD we always have $m_1=m_2$.
The result is
\begin{equation}\label{eq:tc_qqpqpb}
  \begin{split}
  \langle P_{q\to \bar{q}'q'q}(p_1,p_2,p_3)\rangle=&\;
  \frac{C_FT_R}{2}\frac{s_{123}-m_3^2}{s_{12}}\bigg[
    -\frac{t_{12,3}^2}{s_{12}(s_{123}-m_3^2)}+
    \frac{4z_3+(z_1-z_2)^2}{z_1+z_2}-\frac{4m_3^2}{s_{123}-m_3^2}\\
    &\;\qquad\qquad\qquad\qquad+\bigg(1-2\eps+\frac{4m_1^2}{s_{12}}\bigg)
    \bigg(z_1+z_2-\frac{s_{12}}{s_{123}-m_3^2}\bigg)\bigg]\;.
  \end{split}
\end{equation}
The variable $t_{12,i}$ is defined as an extension of Eq.~(22)
in~\cite{Catani:1999ss} to the massive on-shell case, using the fact that
in QCD the final-state parton masses for particles $1$ and $2$ must be
equal.\footnote{Equation~\eqref{eq:def_cg_t123} differs from $t_{12,i}$
  in Ref.~\cite{Dhani:2023uxu}, because our definition is based
  on the diagrammatical origin of the variable.}
\begin{equation}\label{eq:def_cg_t123}
    t_{12,3}=(s_{123}-m_{123}^2)\,S^\mu(p_3,p_{12})\, s_{12}D_\mu(p_1,p_2,\bar{n})=
    2\,\frac{z_1 \tilde{s}_{23}-z_2 \tilde{s}_{13}}{z_1+z_2}+\frac{z_1-z_2}{z_1+z_2}\,s_{12}\;.
\end{equation}
We have also defined the invariant masses $s_{ij}=(p_i+p_j)^2$ and the
scalar products $\tilde{s}_{ij}=2p_ip_j$. The extension of the scalar products
to the three-parton case is given by 
$\tilde{s}_{ijk}=\tilde{s}_{ij}+\tilde{s}_{ik}+\tilde{s}_{jk}$.
Equation~\eqref{eq:tc_qqpqpb} contains a component arising from the scalar
radiator in Eq.~\eqref{eq:tree_level_building_blocks} and the gluon splitting
function in Eq.~\eqref{eq:coll_gqq}. It can be interpreted as a squark-to-quark
splitting function and is given by
\begin{equation}\label{eq:tc_qqpqpb_scalar}
  \begin{split}
  P_{\tilde{q}\to \bar{q}'q'\tilde{q}}(p_1,p_2,p_3)
  =\frac{C_FT_R}{2}\frac{s_{123}-m_3^2}{s_{12}}\left[\frac{4z_3}{z_1+z_2}
    -\frac{4m_3^2}{s_{123}-m_3^2}+\frac{s_{12}}{s_{123}-m_3^2}
    \bigg(1-\frac{t_{12,3}^2}{s_{12}^2}\bigg)\right]\;.
  \end{split}
\end{equation}
The first one-to-three parton splitting function with a non-trivial structure
is the same flavor quark to three quark splitting. There are two contributing
diagrams, shown in Fig.~\ref{fig:one-to-three_splittings_quark}~(b) and~(c).
They lead to two terms of the form of Eq.~\eqref{eq:tc_qqpqpb} and an
interference term
\begin{equation}\label{eq:tc_qqqb}
  \langle P_{q\to \bar{q}qq}(p_1,p_2,p_3)\rangle=
  \Big(\langle P_{q\to \bar{q}'q'q}(p_1,p_2,p_3)\rangle+
  \langle P^{\rm(id)}_{q\to \bar{q}qq}(p_1,p_2,p_3)\rangle\Big)+
  \Big(2\leftrightarrow 3\Big)\;.
\end{equation}
The interference contribution is given in terms of the function
\begin{equation}\label{eq:tc_qqqb_id}
  \begin{split}
  &\langle P^{\rm(id)}_{q\to \bar{q}qq}(p_1,p_2,p_3)\rangle=
  C_F\bigg(C_F-\frac{C_A}{2}\bigg)\bigg\{
  -\frac{(s_{123}-m^2)^2}{s_{12}s_{13}}\frac{z_1}{2}\bigg[\frac{1+z_1^2}{(1-z_2)(1-z_3)}
  -\eps\bigg(1+2\frac{1-z_2}{1-z_3}\bigg)-\eps^2\bigg]\\
  &\qquad+\frac{s_{123}-m^2}{s_{12}}\bigg[\frac{1+z_1^2}{1-z_2}-\frac{2z_2}{1-z_3}
  -\eps\bigg(\frac{(1-z_3)^2}{1-z_2}+1+z_1-\frac{2z_2}{1-z_3}\bigg)-\eps^2(1-z_3)\bigg]
  +(1-\eps)\bigg(\frac{2s_{23}}{s_{12}}-\eps\bigg)\\
  &\qquad+\frac{m^2}{s_{12}s_{13}}\bigg[\,
  s_{23}\,\frac{(1+z_1)(z_1-z_2z_3)}{(1-z_2)(1-z_3)}
  +\eps\Big(z_1(s_{123}-m^2)+s_{23}+6s_{12}\Big)\bigg]\\
  &\qquad-\frac{2m^2}{s_{13}}\frac{(1+z_1)(1+z_2z_3)-(z_2-z_3)^2}{(1-z_2)(1-z_3)}
  +\frac{4m^4}{s_{12}s_{13}}\bigg(\frac{2z_2}{1-z_3}
  -\frac{z_1(1+z_2)}{1-z_2}-\eps\bigg)\bigg\}\;.
  \end{split}
\end{equation}

\subsection{Gluon emission off quarks}
\label{sec:one_to_three_splittings_quark_rad}
\begin{figure}
  \centerline{\includegraphics[scale=0.36]{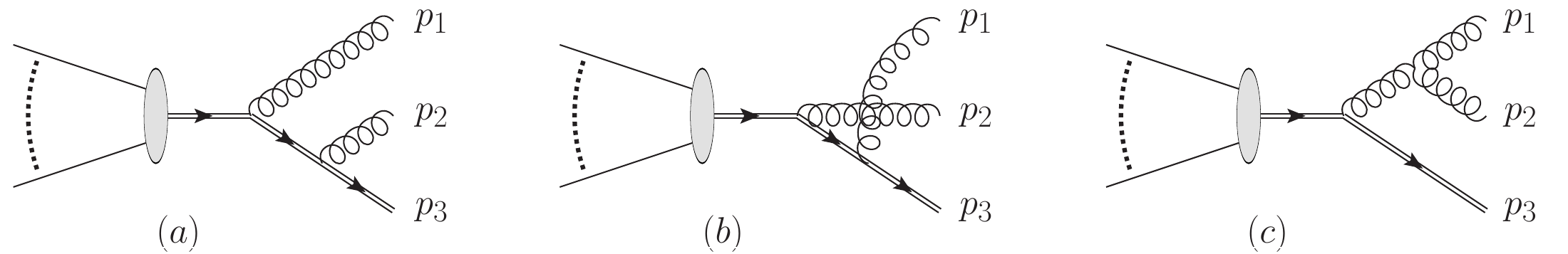}}
  \caption{Feynman diagrams contributing to the one-to-three gluon emission splittings
    with massive quarks, computed in Sec.~\ref{sec:one_to_three_splittings_quark_rad}.
    The double line indicates a quark of mass $m$.
  \label{fig:one-to-three_splittings_quark_gluon}}
\end{figure}
In this section we present the splitting function describing gluon radiation off
massive quarks. Following Ref.~\cite{Catani:1999ss}, we separate the abelian from
the non-abelian component according to
\begin{equation}\label{eq:tc_qgg_all}
  \langle P_{q\to ggq}(p_1,p_2,p_3)\rangle=
  \langle P^{\rm(ab)}_{q\to ggq}(p_1,p_2,p_3)\rangle
  +\langle P^{\rm(nab)}_{q\to ggq}(p_1,p_2,p_3)\rangle\;.
\end{equation}
The relevant Feynman diagrams are shown in
Fig.~\ref{fig:one-to-three_splittings_quark_gluon}. The result for the abelian
component is
\begin{equation}\label{eq:tc_qgg}
  \begin{split}
  &\langle P^{\rm(ab)}_{q\to ggq}(p_1,p_2,p_3)\rangle=C_F^2\,\bigg\{
  \frac{\tilde{s}_{123}^2}{2\tilde{s}_{13}\tilde{s}_{23}}\,z_3\bigg[\frac{1+z_3^2}{z_1z_2}
  -\eps\frac{z_1^2+z_2^2}{z_1z_2}-\eps(1+\eps)
  -\frac{4m^2}{\tilde{s}_{123}}\bigg(\frac{1-z_3}{z_1z_2}
  +(1-\eps)\frac{z_1}{z_3}\bigg)\bigg]\\
  &\;\qquad+\frac{\tilde{s}_{123}}{\tilde{s}_{13}}\bigg[\frac{z_3(1-z_1)+(1-z_2)^3}{z_1z_2}
  +\eps^2(1+z_3)-\eps(z_1^2+z_1z_2+z_2^2)\frac{1-z_2}{z_1z_2}
  -\frac{4m^2}{\tilde{s}_{13}}\bigg(\frac{1-z_2}{z_2}+(1-\eps)\frac{z_2}{2}\bigg)\bigg]\\
  &\;\qquad+\eps(1-\eps)-\frac{\tilde{s}_{23}}{\tilde{s}_{13}}\,(1-\eps)^2+\frac{4m^2}{\tilde{s}_{13}}
  +\frac{1}{2}\bigg(\frac{2m^2}{\tilde{s}_{13}}+\frac{2m^2}{\tilde{s}_{23}}\bigg)^2\,\bigg\}
  +(1\leftrightarrow 2)\;.
  \end{split}
\end{equation}
The non-abelian part of the splitting function is given by
\begin{equation}\label{eq:tc_qgg_nab}
  \begin{split}
  &\langle P^{\rm(nab)}_{q\to ggq}(p_1,p_2,p_3)\rangle=
  -\frac{C_A}{2C_F}P^{\rm(ab)}_{q\to ggq}(p_1,p_2,p_3)+C_FC_A\,\bigg\{
  \frac{1-\eps}{4}\bigg(\frac{t_{12,3}^2}{s_{12}^2}+1\bigg)
  -(1-\eps)^2\frac{\tilde{s}_{23}}{2\tilde{s}_{13}}\\
  &\;\qquad+\frac{\tilde{s}_{123}^2}{2s_{12}\tilde{s}_{13}}\bigg[\frac{(1-z_3)^2(1-\eps)+2z_3}{z_2}
  +\frac{z_2^2(1-\eps)+2(1-z_2)}{1-z_3}-\frac{4m^2}{\tilde{s}_{123}}\bigg(1+\frac{z_1^2}{z_2(1-z_3)}\bigg)\bigg]\\
  &\;\qquad+\frac{\tilde{s}_{123}}{2s_{12}}\bigg[(1-\eps)\frac{z_1(2-2z_1+z_1^2)-z_2(6-6z_2+z_2^2)}{z_2(1-z_3)}
  +2\eps\,\frac{z_3(z_1-2z_2)-z_2}{z_2(1-z_3)}+\frac{4m^2}{\tilde{s}_{123}}\bigg]\\
  &\;\qquad+\frac{\tilde{s}_{123}}{2\tilde{s}_{13}}\bigg[(1-\eps)\frac{(1-z_2)^3+z_3^2-z_2}{z_2(1-z_3)}
  -\eps\,\frac{2(1-z_2)(z_2-z_3)}{z_2(1-z_3)}-\eps(1-\eps)(1-z_1)\\
  &\;\qquad\qquad\qquad-\frac{4m^2}{\tilde{s}_{13}}\bigg(\frac{1-z_2}{z_2}+(1-\eps)\frac{z_2}{2}\bigg)\bigg]
  +\frac{2m^2}{\tilde{s}_{13}}+\frac{2m^4}{\tilde{s}_{13}^2}+(1\leftrightarrow 2)\bigg\}\;.
  \end{split}
\end{equation}

\subsection{Gluon emission off scalars}
\label{sec:one_to_three_splittings_scalar}
The results of Sec.~\ref{sec:one_to_three_splittings_quark_rad}
are the only heavy quark splittings that overlap with the two-gluon scalar dipole
radiator function. The general form of this dipole radiator will be computed in
Sec.~\ref{sec:two-gluon_scalar_radiators}. In order to prepare the discussion of its
physical features, we present here the scalar splitting functions, which contribute
to Eqs.~\eqref{eq:tc_qgg} and~\eqref{eq:tc_qgg_nab}. The abelian component is
\begin{equation}\label{eq:tc_sqgg}
  \begin{split}
  P^{\rm(ab)}_{\tilde{q}\to gg\tilde{q}}(p_1,p_2,p_3)=&\;C_F^2\,\bigg\{
  \frac{\tilde{s}_{123}^2}{\tilde{s}_{13}\tilde{s}_{23}}\bigg[\frac{z_3^2}{z_1z_2}
  -\frac{2m^2}{\tilde{s}_{123}}\frac{z_3(1-z_3)}{z_1z_2}\bigg]
  +\frac{\tilde{s}_{123}}{\tilde{s}_{13}}\bigg[\frac{2z_3(1-z_2)}{z_1 z_2}
  -\frac{4m^2}{\tilde{s}_{13}}\frac{1-z_2}{z_2}\bigg]\\
  &\;\qquad\quad+(1-\eps)+\frac{4m^2}{\tilde{s}_{13}}+\frac{1}{2}
  \bigg(\frac{2m^2}{\tilde{s}_{13}}+\frac{2m^2}{\tilde{s}_{23}}\bigg)^2\bigg\}
  +(1\leftrightarrow 2)\;.
  \end{split}
\end{equation}
In the non-abelian case, we obtain
\begin{equation}\label{eq:tc_sqgg_nab}
  \begin{split}
  P^{\rm(nab)}_{\tilde{q}\to gg\tilde{q}}(p_1,p_2,p_3)=&\;
  -\frac{C_A}{2C_F}P^{\rm(ab)}_{\tilde{q}\to gg\tilde{q}}(p_1,p_2,p_3)
  +C_FC_A\,\bigg\{\frac{1-\eps}{4}\bigg(\frac{t_{12,3}^2}{s_{12}^2}+1\bigg)\\
  &\;+\frac{\tilde{s}_{123}^2}{s_{12}\tilde{s}_{13}}\bigg[\frac{z_3}{z_2}+\frac{1-z_2}{1-z_3}
  -\frac{2m^2}{\tilde{s}_{123}}\frac{z_1^2}{z_2(1-z_3)}\bigg]
  +\frac{\tilde{s}_{123}}{\tilde{s}_{13}}\frac{1-z_2}{z_1}\bigg[\frac{z_3}{z_2}-\frac{1-z_2}{1-z_3}
  -\frac{2m^2}{\tilde{s}_{13}}\frac{z_1}{z_2}\bigg]\\
  &\;+\frac{\tilde{s}_{123}}{s_{12}}\bigg[\frac{z_3}{z_2}
  -\frac{1+3z_3}{1-z_3}+\frac{2m^2}{\tilde{s}_{123}}\bigg]
  -\frac{\tilde{s}_{123}^2}{\tilde{s}_{13}\tilde{s}_{23}}
  \bigg[\frac{m^2}{s_{12}}-\frac{m^2}{\tilde{s}_{123}}\bigg]
  +\frac{2m^2}{\tilde{s}_{13}}+\frac{2m^4}{\tilde{s}_{13}^2}
  +(1\leftrightarrow 2)\bigg\}\;.
  \end{split}
\end{equation}

\subsection{Gluon splitting to quarks}
\label{sec:one_to_three_splittings_gluon}
\begin{figure}
  \centerline{\includegraphics[scale=0.36]{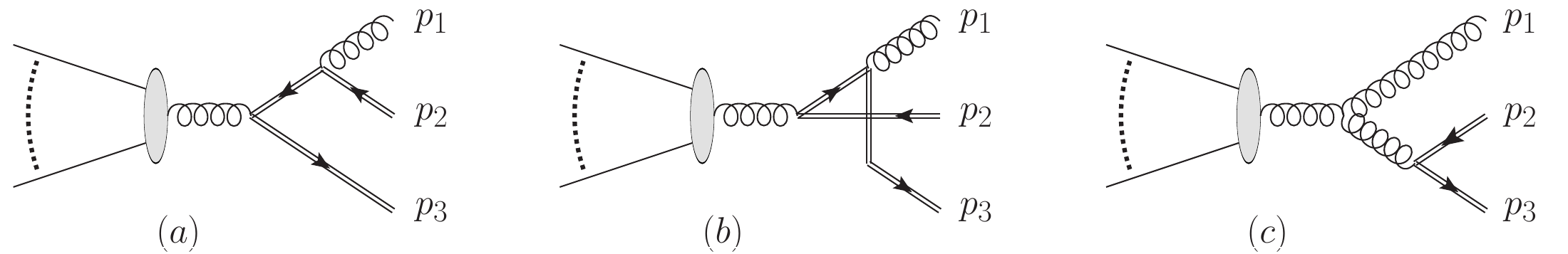}}
  \caption{Feynman diagrams contributing to the one-to-three gluon splitting into
    massive quarks computed in Sec.~\ref{sec:one_to_three_splittings_gluon}.
    The double line indicates a quark of mass $m$.
  \label{fig:one-to-three_splittings_gluon_quark}}
\end{figure}
In this section we quote the gluon splitting functions into a quark antiquark pair
and a gluon. Following the notation in~\cite{Catani:1999ss}, we separate the result
into an abelian and a non-abelian component:
\begin{equation}\label{eq:tc_gqqb}
  P^{\mu\nu}_{g\to gq\bar{q}}(p_1,p_2,p_3)=
  P^{\mu\nu\,\rm(ab)}_{g\to gq\bar{q}}(p_1,p_2,p_3)+
  P^{\mu\nu\,\rm(nab)}_{g\to gq\bar{q}}(p_1,p_2,p_3)\;.
\end{equation}
The Feynman diagrams for this process
are shown in Fig.~\ref{fig:one-to-three_splittings_gluon_quark}.
The abelian component is given by
\begin{equation}\label{eq:tc_gqqb_ab}
  \begin{split}
  &P^{\mu\nu\,\rm(ab)}_{g\to gq\bar{q}}(p_1,p_2,p_3)=C_FT_R\bigg\{
  d^{\mu\nu}(p_{123},\bar{n})\bigg[\,2s_{123}\bigg(\frac{\tilde{s}_{23}}{\tilde{s}_{12}\tilde{s}_{13}}
    -\frac{m^2}{\tilde{s}_{12}^2}-\frac{m^2}{\tilde{s}_{13}^2}\bigg)
    +(1-\eps)\bigg(\frac{\tilde{s}_{12}}{\tilde{s}_{13}}+\frac{\tilde{s}_{13}}{\tilde{s}_{12}}\bigg)-2\eps\,\bigg]\\
  &\;\qquad+\frac{4\tilde{s}_{123}}{\tilde{s}_{12}\tilde{s}_{13}}
  \bigg(\tilde{p}_{2,13}^\mu\tilde{p}_{3,12}^\nu
    +\tilde{p}_{3,12}^\mu\tilde{p}_{2,13}^\nu\bigg)
    +8m^2\bigg(\frac{\tilde{p}_{2,13}^\mu\tilde{p}_{2,13}^\nu}{\tilde{s}_{13}^2}
    +\frac{\tilde{p}_{3,12}^\mu\tilde{p}_{3,12}^\nu}{\tilde{s}_{12}^2}\bigg)
    -(1-\eps)\,\frac{4s_{123}}{\tilde{s}_{12}\tilde{s}_{13}}\,
    \tilde{p}_{1,23}^\mu\,\tilde{p}_{1,23}^\nu
    \bigg\}\;.
  \end{split}
\end{equation}
The non-abelian part is given by
\begin{equation}\label{eq:tc_gqqb_nab}
  \begin{split}
    &P^{\mu\nu\,\rm(nab)}_{g\to gq\bar{q}}(p_1,p_2,p_3)=
    -\frac{C_A}{2C_F}\,P^{\mu\nu\,\rm(ab)}_{g\to gq\bar{q}}(p_1,p_2,p_3)+\frac{C_AT_R}{4}\bigg\{
    \frac{\tilde{s}_{123}}{s_{23}^2}\bigg[-d^{\mu\nu}(p_{123},\bar{n})\frac{t_{23,1}^2}{\tilde{s}_{123}}
    -16\,\frac{1-z_1}{z_1}\,\tilde{p}_{2,3}^\mu\tilde{p}_{2,3}^\nu\,\bigg]\\
    &\;\quad+16m^2\bigg[\frac{\tilde{p}_{2,3}^\mu\tilde{p}_{2,3}^\nu}{z_1}
    \bigg(\!\frac{z_1}{\tilde{s}_{12}^2}-\frac{2(1-z_1)}{s_{23}^2}\bigg)
    -\frac{\tilde{p}^\mu_{1,23}\tilde{p}^\nu_{2,3}+\tilde{p}^\mu_{2,3}\tilde{p}^\nu_{1,23}}{1-z_1}
    \bigg(\!\frac{z_2}{\tilde{s}_{13}^2}+\frac{1-z_1}{\tilde{s}_{13}s_{23}}\!\bigg)
    +\frac{\tilde{p}^\mu_{1,23}\tilde{p}^\nu_{1,23}}{(1-z_1)^2}
    \bigg(\!\frac{z_2}{\tilde{s}_{13}}+\frac{1-z_1}{s_{23}}\!\bigg)^{\!2}\,\bigg]\\
    &\;\quad+d^{\mu\nu}(p_{123},\bar{n}) \bigg[\,\frac{2\tilde{s}_{13}}{\tilde{s}_{12}}(1-\eps)
    +\frac{2\tilde{s}_{123}}{\tilde{s}_{12}}\bigg(\frac{1-z_3}{z_1(1-z_1)}-2\bigg)
    +\frac{2s_{123}}{s_{23}}\frac{1-z_1+2z_1^2}{z_1(1-z_1)}-1-\frac{4m^2s_{123}}{\tilde{s}_{13}^2}\\
    &\;\quad\qquad-\frac{4m^2}{\tilde{s}_{13}}\bigg(2-\frac{1-z_2}{z_1(1-z_1)}\bigg)\,\bigg]
    +\frac{s_{123}}{\tilde{s}_{12}s_{23}} \bigg[\,
    2s_{123}\,d^{\mu\nu}(p_{123},\bar{n})\frac{z_2(1-2z_1)}{z_1(1-z_1)}
    -16\tilde{p}_{3,12}^\mu\tilde{p}_{3,12}^\nu\frac{z_2^2}{z_1(1-z_1)}\\
    &\;\quad\qquad+8(1-\eps)\tilde{p}_{2,13}^\mu\tilde{p}_{2,13}^\nu
    +4(\tilde{p}_{2,13}^\mu \tilde{p}_{3,12}^\nu+\tilde{p}_{3,12}^\mu \tilde{p}_{2,13}^\nu)
    \bigg( \frac{2z_2 (z_3-z_1)}{z_1 (1-z_1)}+1-\eps\bigg)\bigg]+(2\leftrightarrow 3)\bigg\}\;.
  \end{split}
\end{equation}
Here, $\tilde{p}_{i,j}^\mu$ are the generalized transverse momenta defined in Eq.~\eqref{eq:def_z_ptilde}.
They fulfill momentum conservation in the form $\tilde{p}_{i,j}^\mu+\tilde{p}_{j,i}^\mu=0$
for any $i$ and $j$, as well as $\tilde{p}_{1,23}^\mu+\tilde{p}_{2,13}^\mu+\tilde{p}_{3,12}^\mu=0$.
We have not listed explicitly $\bar{n}$-dependent components of the splitting functions,
as they vanish when multiplied by the polarization tensor of the intermediate gluon.

\section{Scalar radiator functions}
\label{sec:scalar_radiators}
\begin{figure}
  \centerline{\includegraphics[scale=0.36]{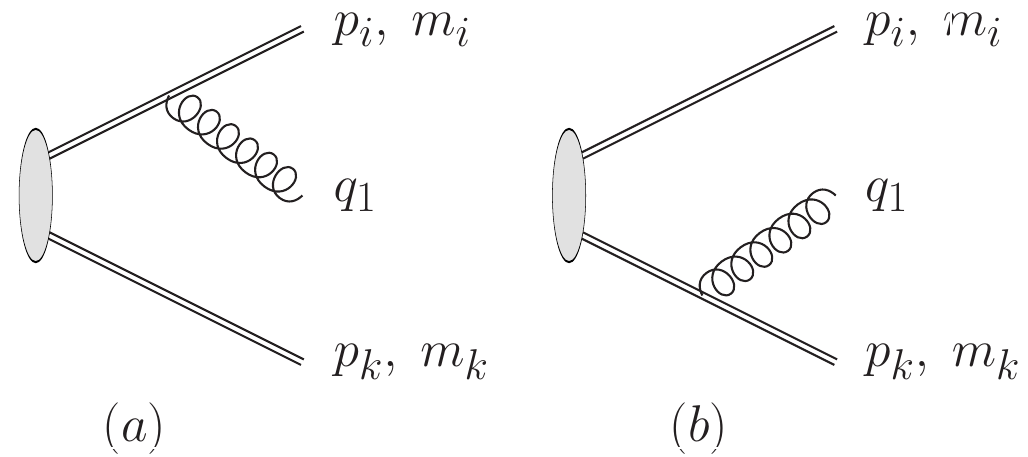}}
  \caption{Feynman diagrams contributing to the scalar radiator function
    for a single gluon emission. The double lines indicate quark masses
    of the scalar particles.
  \label{fig:scalar_single_emission}}
\end{figure}
In this section we compute the scalar radiator functions, which describe the
production of one or more partons through coherent radiation off a QCD 
dipole antenna. In contrast to the splitting functions in Sec.~\ref{sec:method}
and~\ref{sec:three-parton_tree-level}, they account for the precise direction
of flight of the color anti-charge, which must accompany the initial state 
charged parton in a splitting configuration such as
Fig.~\ref{fig:one-to-two_splittings}~(a) or
Fig.~\ref{fig:one-to-three_splittings_quark_gluon} due to color conservation.

The simplest such function is the one-gluon radiator shown in
Fig.~\ref{fig:scalar_single_emission}. Using the formalism and 
notation of~\cite{Bassetto:1984ik,Catani:1996vz} generalized to
the scalar QCD case~\cite{Campbell:2025lrs}, the dipole approximation
to the complete radiator function for a single gluon can be written
in terms of color insertion operators, ${\bf T}$, and a space-time
dependent part, $\mathcal{S}_{i;k}$, as follows
\begin{equation}\label{eq:one-gluon_insertion_1}
  \begin{split}
  \mathcal{S}_g(\{p\};q_1;\bar{n}) = &\; \sum_{i,k}
    \hat{\bf T}_i\hat{\bf T}_k\,
    \mathcal{S}_{i;k}(q_1;\bar{n})\;.
  \end{split}
\end{equation}
For massive emitters and an on-shell gluon with momentum $q_1$,
the function $\mathcal{S}_{i;k}$ is given by
\begin{equation}\label{eq:one-gluon_massive_radiator}
  \begin{split}
    \mathcal{S}_{i;k}(q_1;\bar{n})=&\;
    \frac{1}{p_{i1}^2-m_i^2}\frac{2z_i}{z_1}\bigg(1-\frac{p_k^2-m_k^2}{p_{k1}^2-m_k^2}\bigg)
    +\frac{1}{p_{k1}^2-m_k^2}\frac{2z_k}{z_1}\bigg(1-\frac{p_i^2-m_i^2}{p_{i1}^2-m_i^2}\bigg)
    -\frac{4p_ip_k}{(p_{i1}^2-m_i^2)(p_{k1}^2-m_k^2)}\;.
  \end{split}
\end{equation}
In order to simplify the notation in Sec.~\ref{sec:decomposition}, we
also introduce a compact notation for the gluon-spin dependent radiator
function. It is given by
\begin{equation}\label{eq:one-gluon_insertion_spin}
  \begin{split}
  \mathcal{S}^{\mu\nu}_g(\{p\};q_1)=&\;
  \sum_{i,k}
    \hat{\bf T}_i\hat{\bf T}_k\,
    \mathcal{S}^{\mu\nu}_{i;k}(q_1)\;,
    \quad&&\text{where}\quad
    &\mathcal{S}^{\mu\nu}_{i;k}(q_1)=&\;
    S^\mu(p_i,q_1)S^\nu(p_k,q_1)\;.
  \end{split}
\end{equation}
For the decomposition of the two-gluon splitting functions in
Sec.~\ref{sec:decomposition}, we will need the off-shell
radiator~\cite{Campbell:2025lrs} for massive particles. It is determined by
using the momentum shift in Eq.~\eqref{eq:momentum_shift_massive} to define
the external wave function of the gluon. Its space-time dependent part reads
\begin{equation}\label{eq:one-gluon_shifted_massive_radiator}
  \begin{split}
  \bar{\mathcal{S}}_{i;k}(q_1;\bar{n})
  =\frac{1}{p_{i1}^2-m_i^2}\frac{2z_i}{z_1}
  \bigg(1-\frac{p_k^2-m_k^2}{p_{k1}^2-m_k^2}-\frac{q_1^2}{p_{k1}^2-m_k^2}\frac{z_{k1}}{z_1}\bigg)
  -\frac{2p_ip_k}{(p_{i1}^2-m_i^2)(p_{k1}^2-m_k^2)}+(i\leftrightarrow k)\;.
  \end{split}
\end{equation}
The squark splitting function in Eq.~\eqref{eq:coll_q_to_qg_components}
is a special case of this radiator, $P_{\tilde{q}\to\tilde{q}}(p_1,p_2)=
C_F\,(p_{12}^2-m_1^2)\,\bar{\mathcal{S}}_{1;1}(p_2;\bar{n})/2\;$.

\subsection{Quark-antiquark radiator}
\label{sec:two-quark_scalar_radiators}
\begin{figure}
  \centerline{\includegraphics[scale=0.36]{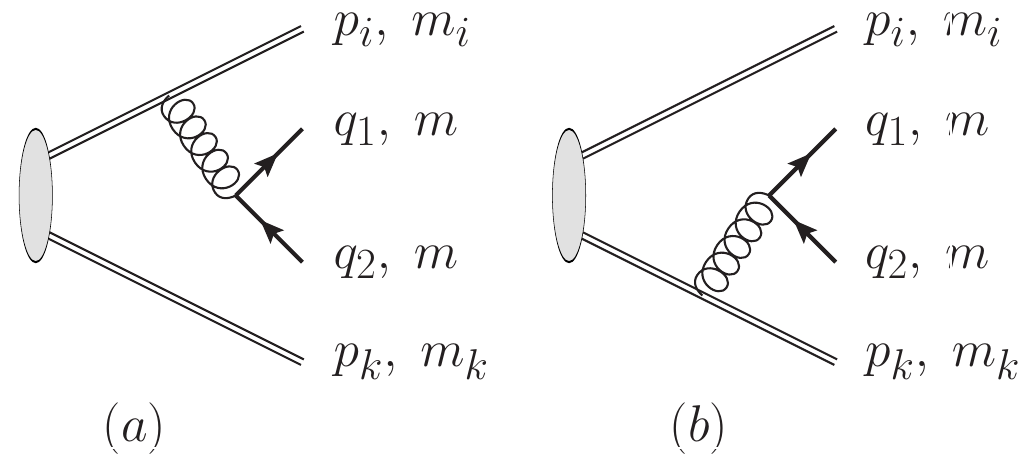}}
  \caption{Feynman diagrams contributing to the scalar radiator for a quark-antiquark pair
    computed in Sec.~\ref{sec:two-quark_scalar_radiators}.
    The double line indicates quark masses of the scalar particles,
    while the thick solid line indicates the masses of the emitted quarks.
  \label{fig:two-quark_scalar_radiators}}
\end{figure}
The scalar dipole approximation to radiation of a quark-antiquark pair is obtained from
the product of the spin-dependent one-gluon radiator in Eq.~\eqref{eq:one-gluon_insertion_spin}
and the gluon-to-quark splitting tensor in Eq.~\eqref{eq:coll_gqq}. The relevant
Feynman diagrams are shown in Fig.~\ref{fig:two-quark_scalar_radiators}. Due to the
factorized form of the result, the color decomposition is identical to the
single gluon emission case in Eq.~\eqref{eq:one-gluon_insertion_spin}. It reads
\begin{equation}\label{eq:scalar_emission_quark_pair_2}
  \begin{split}
  \mathcal{S}_{q\bar{q}}(\{p\};q_1,q_2;\bar{n}) = &\; \sum_{i,k}
    \hat{\bf T}_i\hat{\bf T}_k\,T_R\,\mathcal{S}^{(q\bar{q})}_{i;k}(q_1,q_2;\bar{n})\;,
  \end{split}
\end{equation}
The space-time dependent component is given by
\begin{equation}\label{eq:scalar_emission_quark_pair_individual}
  \begin{split}
  \mathcal{S}^{(q\bar{q})}_{i;k}(q_1,q_2;\bar{n}) = &\;
  \frac{2}{(s_{i12}-m_i^2)(s_{k12}-m_k^2)}\bigg[\,
    \frac{2}{s_{12}}\bigg(
    \frac{z_i(s_{k12}-m_k^2)+z_k(s_{i12}-m_i^2)}{z_1+z_2}
    -\tilde{s}_{ik}\bigg)+1-\frac{t_{12,i}t_{12,k}}{s_{12}^2}\,\bigg]\;.
  \end{split}
\end{equation}
For $i=k$, this result is proportional to the scalar splitting function in
Eq.~\eqref{eq:tc_qqpqpb_scalar}, cf.\ Eq.~\eqref{eq:split_qbpqpq_assembly}.

\subsection{Two-gluon radiator}
\label{sec:two-gluon_scalar_radiators}
\begin{figure}
  \centerline{\includegraphics[scale=0.36]{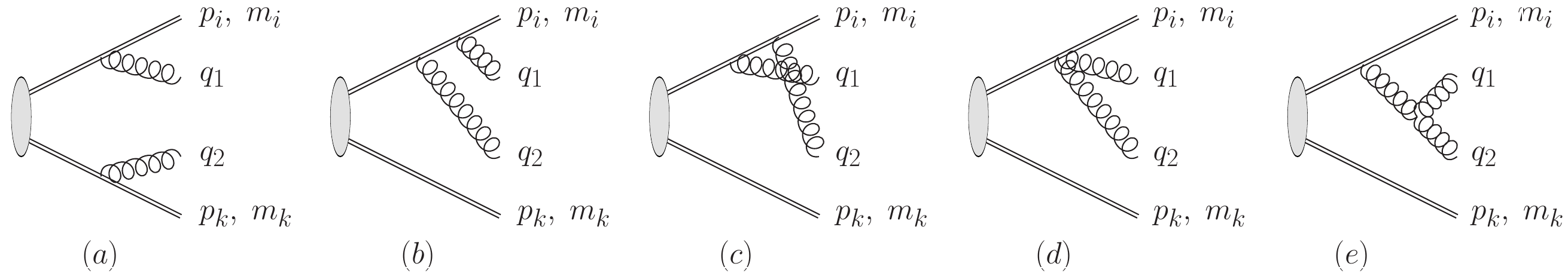}}
  \caption{Feynman diagrams contributing to the scalar radiator for
    two gluons, computed in Sec.~\ref{sec:two-gluon_scalar_radiators}.
    The double line indicates quark masses of the scalar particles.
  \label{fig:two-gluon_scalar_radiators}}
\end{figure}

The dipole approximation to the two-gluon radiator function has a more complicated
structure. The relevant Feynman diagrams are shown in Fig.~\ref{fig:two-gluon_scalar_radiators}.
To understand its form, it is helpful to write the result in terms of the
two-gluon current~\cite{Catani:1999ss}. It is interesting to note that this current
is only minimally modified compared to its massless counterpart. For an on-shell
radiator we find
\begin{equation}\label{eq:two-gluon_current}
    \begin{split}
        &J^{\mu\nu}_{ab}(q_1,q_2)=
        \sum_{i,k}\big\{\hat{\bf T}^a_i,\hat{\bf T}^b_k\big\}\,
          \mathcal{J}^{{\rm(ab)},\mu\nu}_{ik}(q_1,q_2)
        +\sum_i\,if^{abc}\hat{\bf T}_i^c\,
          \mathcal{J}^{{\rm(nab)},\mu\nu}_{i}(q_1,q_2)\;,
    \end{split}
\end{equation}
with the color-stripped abelian and non-abelian components
\begin{equation}\label{eq:two-gluon_currents}
  \begin{split}
    &\mathcal{J}^{{\rm(ab)},\mu\nu}_{ik}(q_1,q_2)=
    \frac{1}{2}\,S^\mu(p_i,q_1)S^\nu(p_k,q_2)\\
    &\;\qquad+\frac{\delta_{ik}}{(p_i+q_{12})^2-m_i^2}\,\bigg[\, 
    q_1^\nu S^\mu(p_i,q_1)+q_2^\mu S^\nu(p_i,q_2)
    -q_1q_2\,S^\mu(p_i,q_1)S^\nu(p_i,q_2)-g^{\mu\nu}\bigg]\;,\\
    &\mathcal{J}^{{\rm(nab)},\mu\nu}_{i}(q_1,q_2)=
    S_\rho(p_i,q_{12})
    \Big(d^{\rho\nu}(q_{12})S^\mu(q_2,q_1)-d^{\rho\mu}(q_{12})S^\nu(q_1,q_2)\Big)\\
    &\;\qquad+\frac{1}{(p_i+q_{12})^2-m_i^2}\,
    \bigg[\,q_1^\nu S^\mu(p_i,q_1)-q_2^\mu S^\nu(p_i,q_2)
    +p_i(q_2-q_1)\,S^\mu(p_i,q_1)S^\nu(p_i,q_2)
    +\frac{t_{12,i}}{q_{12}^2}\,g^{\mu\nu}\,\bigg]\;.
  \end{split}
\end{equation}
Changes compared to the massless result only occur due to the definition
of $t_{12,i}$ (cf. Eq.~\eqref{eq:def_cg_t123}), and due to the difference
in the propagators.
The squared two-gluon current is given by~\cite{Campbell:2025lrs}
\begin{equation}\label{eq:squared_two-gluon_current}
  \begin{split}
    &\big[J^{ab}_{\mu\nu}(q_1,q_2)\big]^\dagger
    d^{\mu\rho}(q_1,\bar{n})d^{\nu\sigma}(q_2,\bar{n})
    J^{ab}_{\rho\sigma}(q_1,q_2)=2\sum_{i,k}\sum_{l,m}
    \Big\{\hat{\bf T}^a_i\hat{\bf T}^a_l,
    \hat{\bf T}^b_k\hat{\bf T}^b_m\Big\}\,
    \mathcal{S}^{\rm(ab)}_{i,k;l,m}(q_1,q_2;\bar{n})\\
    &\;\quad+2\sum_{i,k}\sum_{l}
    \Big(\Big\{\hat{\bf T}^a_i\hat{\bf T}^a_l,
    \hat{\bf T}^b_k\hat{\bf T}^b_l\Big\}+
    \Big\{\hat{\bf T}^a_l\hat{\bf T}^a_i,
    \hat{\bf T}^b_l\hat{\bf T}^b_k\Big\}\Big)\,
    \mathcal{S}^{\rm(ab)}_{i,k;l}(q_1,q_2;\bar{n})
    +2\sum_{i,l}\Big\{\hat{\bf T}^a_i\hat{\bf T}^a_l,
    \hat{\bf T}^b_i\hat{\bf T}^b_l\Big\}\,
    \mathcal{S}^{\rm(ab)}_{i;l}(q_1,q_2)\\
    &\;\quad-\sum_{i,l}\,C_A\,\hat{\bf T}^c_i\hat{\bf T}^c_l\,
    \Big[\;\mathcal{S}^{\rm(nab)}_{i;l}(q_1,q_2;\bar{n})
    -(1-2\delta_{il})\,\mathcal{S}^{\rm(ab)}_{i;l}(q_1,q_2)
    -\mathcal{S}^{\rm(ab)}_{i,i;l}(q_1,q_2;\bar{n})
    -\mathcal{S}^{\rm(ab)}_{l,l;i}(q_1,q_2;\bar{n})\\
    &\;\qquad\qquad\qquad\qquad\qquad
    +\mathcal{S}^{\rm(ab)}_{i,l;i}(q_1,q_2;\bar{n})
    +\mathcal{S}^{\rm(ab)}_{l,i;i}(q_1,q_2;\bar{n})
    +\mathcal{S}^{\rm(ab)}_{i,l;l}(q_1,q_2;\bar{n})
    +\mathcal{S}^{\rm(ab)}_{l,i;l}(q_1,q_2;\bar{n})\,\Big]\;,
  \end{split}
\end{equation}
where we have introduced the abelian and non-abelian radiator functions,
$\mathcal{S}(q_1,q_2;\bar{n})$. For massive on-shell particles they are
defined as an extension of the results in~\cite{Campbell:2025lrs}.
\begin{equation}\label{eq:squared_two-gluon_current_ab}
  \begin{split}
    \mathcal{S}^{\rm(ab)}_{i,k;l,m}(q_1,q_2;\bar{n})=&\;
    \frac{1}{4}\,\mathcal{S}_{i;l}(q_1;\bar{n})
    \mathcal{S}_{k;m}(q_2;\bar{n})\;,\\
    \mathcal{S}^{\rm(ab)}_{i,k;l}(q_1,q_2;\bar{n})=&\;
    \frac{1}{\tilde{s}_{l12}}\frac{s_{il}s_{kl}}{\tilde{s}_{i1}\tilde{s}_{k2}}
    \bigg(\frac{\tilde{s}_{k1}}{s_{kl}\tilde{s}_{l1}}
    +\frac{\tilde{s}_{i2}}{s_{il}\tilde{s}_{l2}}
    -\frac{s_{12}}{\tilde{s}_{l1}\tilde{s}_{l2}}\bigg)\\
    &\hspace*{-15mm}+\frac{1}{\tilde{s}_{l12}}\bigg[\frac{z_l}{z_2}\bigg(
    \frac{\tilde{s}_{il}s_{12}-\tilde{s}_{i2}\tilde{s}_{l1}}{
    \tilde{s}_{i1}\tilde{s}_{l1}\tilde{s}_{l2}}
    -\frac{z_ls_{12}}{2z_1\,\tilde{s}_{l1}\tilde{s}_{l2}}\bigg)
    +\frac{z_i\tilde{s}_{l1}+z_l\tilde{s}_{i1}-z_1\tilde{s}_{il}}{
      z_2\,\tilde{s}_{i1}\tilde{s}_{l1}}\\
    &\hspace*{-15mm}\qquad\qquad-\frac{\tilde{s}_{ik}}{2\tilde{s}_{i1}\tilde{s}_{k2}}
    \bigg(1+\frac{m_k^2+m_l^2}{\tilde{s}_{ik}}
    \frac{2\tilde{s}_{i2}}{\tilde{s}_{l2}}\bigg)
    +\frac{s_{12}(s_{il}+\tilde{s}_{il})(m_k^2+m_l^2)}{
      2\tilde{s}_{i1}\tilde{s}_{k2}\tilde{s}_{l1}\tilde{s}_{l2}}
    +\bigg(\!\!\begin{array}{r@{\,}l}i&\leftrightarrow k\\
    1&\leftrightarrow 2\end{array}\!\!\bigg)\bigg]\;,\\
    \mathcal{S}^{\rm(ab)}_{i;k}(q_1,q_2)=&\;
    \frac{(\tilde{s}_{i1}\tilde{s}_{k2}-\tilde{s}_{i2}\tilde{s}_{k1})^2
    -2s_{12}\tilde{s}_{ik}(\tilde{s}_{i1}\tilde{s}_{k2}+\tilde{s}_{i2}\tilde{s}_{k1})
    +s_{12}^2\tilde{s}_{ik}^2}{\tilde{s}_{i12}\tilde{s}_{k12}\,
    \tilde{s}_{i1}\tilde{s}_{k1}\tilde{s}_{i2}\tilde{s}_{k2}}\\
    &\;+\frac{2s_{12}}{\tilde{s}_{i12}\tilde{s}_{k12}}
    \bigg(\frac{m_i^2}{\tilde{s}_{i1}\tilde{s}_{i2}}
    +\frac{m_k^2}{\tilde{s}_{k1}\tilde{s}_{k2}}\bigg)
    +\frac{2(1-\eps)}{\tilde{s}_{i12}\tilde{s}_{k12}}\;,
  \end{split}
\end{equation}
and
\begin{equation}\label{eq:squared_two-gluon_current_nab}
  \begin{split}
    &\mathcal{S}^{\rm(nab)}_{i;k}(q_1,q_2;\bar{n})=
    \mathcal{S}^{({\rm nab},p)}_{i;k}(q_1,q_2;\bar{n})
    -2(1-\eps)\,\mathcal{S}^{\mu\nu}_{i;k}(q_{12};\bar{n})
    D_\mu(q_1,q_2,\bar{n})D_\nu(q_1,q_2,\bar{n})\\
    &\;\quad+\frac{1}{4}\,\Big(\bar{\mathcal{S}}_{i;i}(q_{12};\bar{n})
    +\bar{\mathcal{S}}_{k;k}(q_{12};\bar{n})-2\bar{\mathcal{S}}_{i;k}(q_{12};\bar{n})\Big)
    \bigg[\,\mathcal{S}_{i;i}(q_2;\bar{n})+\mathcal{S}_{1;1}(q_2;\bar{n})-2\mathcal{S}_{i;1}(q_2;\bar{n})\\
    &\;\qquad\qquad\qquad-\frac{1}{2}\Big(\mathcal{S}_{i;i}(q_2;\bar{n})+\mathcal{S}_{k;k}(q_2;\bar{n})
    -2\mathcal{S}_{i;k}(q_2;\bar{n})\Big)
    +\big(i\leftrightarrow k\big)+\big(1\leftrightarrow 2\big)
    +\bigg(\!\!\begin{array}{r@{\,}l}i&\leftrightarrow k\\
    1&\leftrightarrow 2\end{array}\!\!\bigg)\bigg]\;,
  \end{split}
\end{equation}
where
\begin{equation}\label{eq:squared_two-gluon_current_pnab}
  \begin{split}
    &\mathcal{S}^{({\rm nab},p)}_{i;k}(q_1,q_2;\bar{n})=
    \bigg[\frac{3\tilde{s}_{ik}}{4\tilde{s}_{i1}\tilde{s}_{k1}}+\frac{2s_{12}}{z_{12}^2}
    \bigg(\frac{z_i}{\tilde{s}_{i12}}-\frac{z_k}{\tilde{s}_{k12}}\bigg)
    \bigg(\frac{z_{i12}}{\tilde{s}_{i12}}-\frac{z_{k12}}{\tilde{s}_{k12}}\bigg)\bigg]
    \bigg(\frac{2\tilde{s}_{i1}}{s_{12}\tilde{s}_{i2}}
    -\frac{\tilde{s}_{ik}}{\tilde{s}_{i2}\tilde{s}_{k2}}\bigg)\\
    &\;\quad+\frac{1}{\tilde{s}_{i12}\tilde{s}_{k12}}\bigg\{
    \frac{3}{2}+\frac{3}{2}\frac{\tilde{s}_{ik}^2}{\tilde{s}_{i1}\tilde{s}_{k1}}
    +\frac{\tilde{s}_{ik}}{2s_{12}}-\frac{z_2}{z_1}
    +\frac{\tilde{s}_{k1}\tilde{s}_{i2}-\tilde{s}_{ik}s_{12}}{
      2\tilde{s}_{i1}\tilde{s}_{k2}}\bigg(1-\frac{\tilde{s}_{ik}}{s_{12}}\bigg)
    +\frac{\tilde{s}_{ik}}{\tilde{s}_{i1}}\bigg(1-\frac{3\tilde{s}_{i12}}{s_{12}}\bigg)\\
    &\;\quad\qquad-\frac{\tilde{s}_{i2}}{\tilde{s}_{i1}}\frac{z_i}{z_2}\bigg(2+
    \frac{z_1}{z_i}+\frac{\tilde{s}_{i12}}{\tilde{s}_{i2}}+\frac{\tilde{s}_{k12}}{\tilde{s}_{i2}}\bigg)
    +\frac{2\tilde{s}_{k1}}{s_{12}}\bigg(\frac{z_1-z_2}{z_1}\frac{z_2-z_i}{z_2}
    -\frac{z_i\tilde{s}_{i12}}{z_2\tilde{s}_{i1}}+\frac{\tilde{s}_{i2}}{
      \tilde{s}_{i1}}-\frac{\tilde{s}_{i1}}{\tilde{s}_{i2}}\frac{2z_i+z_2}{z_1}\bigg)\\
    &\;\quad\qquad-\frac{z_{12}^2}{2z_1z_2}\frac{t_{12,i}t_{12,k}}{s_{12}^2}
    -\frac{t_{12,k}}{2s_{12}}\bigg[\frac{\tilde{s}_{i1}}{\tilde{s}_{i2}}\frac{z_{i2}}{z_1}
      \bigg(1-\frac{z_is_{12}}{z_2\tilde{s}_{i1}}\bigg)
      +\frac{2z_{12}}{z_1}\bigg(1+\frac{\tilde{s}_{i1}}{s_{12}}\bigg)+\frac{3z_i}{z_2}
    -\big(1\!\leftrightarrow\! 2\big)\bigg]\\
    &\;\quad\qquad-\frac{m_i^2}{\tilde{s}_{i1}}\bigg[\frac{2t_{12,k}}{s_{12}}
    -\frac{2s_{ik}s_{12}}{\tilde{s}_{i2}^2}\bigg(1+\frac{s_{12}}{\tilde{s}_{k1}}
    +\frac{\tilde{s}_{i2}\tilde{s}_{k1}}{\tilde{s}_{i1}\tilde{s}_{k2}}\bigg)\bigg]\bigg\}
    +\frac{z_i^2}{2\tilde{s}_{i1}\tilde{s}_{i2}z_1z_2}\bigg(1+
    \frac{\tilde{s}_{i1}-\tilde{s}_{i2}}{\tilde{s}_{i12}}
    \frac{\tilde{s}_{k1}-\tilde{s}_{k2}}{\tilde{s}_{k12}}\bigg)\\
    &\;\quad+\frac{4m_i^2}{\tilde{s}_{i12}^2}
    \frac{\tilde{s}_{i1}\tilde{s}_{k2}+\tilde{s}_{i2}\tilde{s}_{k1}
    -\tilde{s}_{ik}s_{12}}{s_{12}\tilde{s}_{i1}\tilde{s}_{k1}}
    +\frac{m_i^2}{\tilde{s}_{i1}\tilde{s}_{i12}}\bigg[
    \frac{4}{\tilde{s}_{k2}}\bigg(\frac{s_{ik}}{\tilde{s}_{i2}}-\frac{\tilde{s}_{k1}}{s_{12}}
    +\frac{\tilde{s}_{k2}}{s_{12}}\frac{z_1}{z_2}-\frac{\tilde{s}_{k1}}{2\tilde{s}_{i1}}\bigg)
    -\frac{1}{\tilde{s}_{i2}}\frac{2z_i}{z_1}+\frac{1}{\tilde{s}_{i1}}\frac{2z_{i1}}{z_2}\\
    &\;\qquad\quad-\frac{2\tilde{s}_{i1}+s_{12}}{\tilde{s}_{i1}}
    \bigg(\frac{m_i^2}{\tilde{s}_{i2}^2}+\frac{2(m_i^2+m_k^2)}{\tilde{s}_{i2}\tilde{s}_{k2}}\bigg)\bigg]
    +\big(i\leftrightarrow k\big)+\big(1\leftrightarrow 2\big)
    +\bigg(\!\!\begin{array}{r@{\,}l}i&\leftrightarrow k\\
    1&\leftrightarrow 2\end{array}\!\! \bigg)\;.
  \end{split}
\end{equation}
As in the massless parton case, the radiators define the splitting
functions in Eqs.~\eqref{eq:tc_sqgg} and~\eqref{eq:tc_sqgg_nab}. We find
\begin{equation}\label{eq:scalar_tc_correspondence}
  \begin{split}
    P^{\rm(ab)}_{\tilde{q}\to gg\tilde{q}}(p_1,p_2,p_i)=&\;
    (s_{i12}-m_i^2)^2\,C_F^2\Big(\mathcal{S}^{\rm(ab)}_{i,i;i,i}(p_1,p_2;\bar{n})
    +2\mathcal{S}^{\rm(ab)}_{i,i;i}(p_1,p_2;\bar{n})+
    \mathcal{S}^{\rm(ab)}_{i;i}(p_1,p_2)\Big)\;,\\ 
    P^{\rm(nab)}_{\tilde{q}\to gg\tilde{q}}(p_1,p_2,p_i)=&\;
    -(s_{i12}-m_i^2)^2\,\frac{C_FC_A}{4}\,\mathcal{S}^{\rm(nab)}_{i;i}(p_1,p_2;\bar{n})
    -\frac{C_A}{4C_F}P^{\rm(ab)}_{\tilde{q}\to gg\tilde{q}}(p_1,p_2,p_i)\;.
  \end{split}
\end{equation}
In the double-soft gluon limit, Eq.~\eqref{eq:squared_two-gluon_current_nab}
reduces to the massive double-soft eikonal computed in
App.~B of Ref.~\cite{Czakon:2011ve} (see also~\cite{Catani:2019nqv}).

\section{Composition of one-to-three splitting functions}
\label{sec:decomposition}
In this section we address the main question of this work, namely how
the one-to-three parton splitting functions can be decomposed in terms
of the scalar dipole radiators derived in Sec.~\ref{sec:scalar_radiators}
and the lower-order splitting functions in Sec.~\ref{sec:method}.
The results we will derive are structurally identical to the massless case,
which was determined based on the Feynman diagrams that enter the calculations.
This provides a strong cross-check of our technique, as the remainder
functions we will obtain are required to reduce to their massless equivalent
in the case of massless external partons.\footnote{No limiting procedure is
involved when determining the massless results from the massive ones,
the masses can simply be set to zero.} We will make this structure apparent
in all our results.

\subsection{All-quark splitting functions}
\label{sec:decomposition_quark}
We begin with the simplest case, which is the all-quark splitting function
with distinct flavor quarks. Only the diagram
in Fig.~\ref{fig:one-to-three_splittings_quark}~(a) contributes to this
result, therefore the splitting function admits a trivial decomposition,
which is given by
\begin{equation}\label{eq:split_qbpqpq_assembly}
  \begin{split}
    \langle P_{q\to \bar{q}'q'q}(p_1,p_2,p_3)\rangle
    =&\;\frac{s_{123}-m_3^2}{s_{12}}\bigg[\,
    \frac{C_F}{2}\,(s_{123}-m_3^2)\,\mathcal{S}^{\mu\nu}_{3;3}(p_{12})
    +\langle P_{q\to q}^{{\rm(f)}\mu\nu}(p_3,p_{12})\rangle\bigg]
    P_{g\to q,\mu\nu}(p_1,p_2)\;.
  \end{split}
\end{equation}
We note that the scalar contribution can be written in terms of the
two-quark scalar radiator in Eq.~\eqref{eq:scalar_emission_quark_pair_individual},
or in terms of the scalar splitting function, Eq.~\eqref{eq:tc_qqpqpb_scalar}:
\begin{equation}\label{eq:decomposition_qqq}
  P_{\tilde{q}\to \bar{q}'q'\tilde{q}}(p_1,p_2,p_3)=
  \frac{C_F}{2}\frac{(s_{123}-m_3^2)^2}{s_{12}}\,
  \mathcal{S}^{\mu\nu}_{3;3}(p_{12})P_{g\to q,\mu\nu}(p_1,p_2)
  =\frac{C_FT_R}{4}\,(s_{123}-m_3^2)^2\,
  \mathcal{S}^{(q\bar{q})}_{3;3}(p_1,p_2;\bar{n})\;.
\end{equation}
For the all-quark splitting function with same flavor quarks, we obtain
\begin{equation}\label{eq:split_qbqq_assembly}
  \begin{split}
    \langle P_{q\to \bar{q}qq}(p_1,p_2,p_3)\rangle
    =&\;\bigg\{\frac{s_{123}-m^2}{s_{12}} \bigg[
    \,\frac{C_F}{2}\,(s_{123}-m^2)\,\mathcal{S}^{\mu\nu}_{3;3}(p_{12})
    +\langle P_{q\to q}^{{\rm(f)}\mu\nu}(p_3,p_{12})\rangle\bigg]
    P_{g\to q,\mu\nu}(p_1,p_2)\\
    &\qquad\quad+\Big(2\leftrightarrow 3\Big)\,\bigg\}
    +\langle P^{(p)}_{q\to \bar{q}qq}(p_1,p_2,p_3)\rangle\;.
  \end{split}
\end{equation}
The pure splitting contribution is a consequence of interference between
the diagrams in Fig.~\ref{fig:one-to-three_splittings_quark}~(b) and~(c),
and is given in terms of Eq.~\eqref{eq:tc_qqqb_id} as
\begin{equation}\label{eq:pure_split_qbqq}
  \begin{split}
    \langle P^{(p)}_{q\to \bar{q}qq}(p_1,p_2,p_3)\rangle=&\;
    \langle P^{\rm(id)}_{q\to \bar{q}qq}(p_1,p_2,p_3)\rangle+
    \big(2\leftrightarrow 3\big)\;.
  \end{split}
\end{equation}

\subsection{Gluon emission off quarks}
\label{sec:decomposition_quark_gluon}
The quark to quark gluon splitting function overlaps with the scalar dipole
radiator for two-gluon emission in Sec.~\ref{sec:two-gluon_scalar_radiators}.
It allows the following decomposition~\cite{Campbell:2025lrs}
\begin{equation}\label{eq:split_qgg_ab_assembly}
  \begin{split}
    \langle P_{q\to ggq}^{\rm(ab)}&(p_1,p_2,p_3)\rangle
    =\bigg[\,\frac{C_F^2}{4}\,\tilde{s}_{123}^2\,
    \mathcal{S}_{13;13}(p_2;\bar{n})\mathcal{S}_{3;3}(p_1;\bar{n})
    +\frac{C_F}{2}\frac{\tilde{s}_{123}^2}{\tilde{s}_{13}}\,
    \mathcal{S}_{13;13}(p_2;\bar{n})\langle P^{\rm(f)}_{q\to q}(p_3,p_1)\rangle\\
    &\quad\;+\frac{C_F}{2}\,\tilde{s}_{123}\,
    \langle P^{\rm(f)}_{q\to q}(p_{13},p_2)\rangle\,
    \mathcal{S}_{3;3}(p_1;\bar{n})
    +\frac{\tilde{s}_{123}}{\tilde{s}_{13}}\langle P^{\rm(f)}_{q\to q}(p_{13},p_2)\rangle
    \langle P^{\rm(f)}_{q\to q}(p_3,p_1)\rangle
    +\Big(1\leftrightarrow 2\Big)\,\bigg]\\
    &\;+P^{({\rm ab},p)}_{\tilde{q}\to gg\tilde{q}}(p_1,p_2,p_3)
    +\langle P^{({\rm ab},p,{\rm f})}_{q\to ggq}(p_1,p_2,p_3)\rangle\;.
  \end{split}
\end{equation}
where the pure scalar and fermionic components are given by
\begin{equation}\label{eq:pure_split_ab_sqgg}
  \begin{split}
  P^{({\rm ab},p)}_{\tilde{q}\to gg\tilde{q}}(p_1,p_2,p_3)=&\;
  P^{({\rm ab})}_{\tilde{q}\to gg\tilde{q}}(p_1,p_2,p_3)
  -\frac{C_F^2}{4}\,\tilde{s}_{123}^2\,
    \Big[\mathcal{S}_{13;13}(p_2;\bar{n})\mathcal{S}_{3;3}(p_1;\bar{n})
    +\mathcal{S}_{23;23}(p_1;\bar{n})\mathcal{S}_{3;3}(p_2;\bar{n})\Big]\\
  =&\;C_F^2\,\bigg\{
  \frac{\tilde{s}_{123}^2}{\tilde{s}_{13}\tilde{s}_{23}}\bigg[\frac{z_3^2}{z_1z_2}
  -\frac{2m^2}{\tilde{s}_{123}}\frac{z_3(1-z_3)}{z_1z_2}\bigg]
  -\frac{\tilde{s}_{123}}{\tilde{s}_{13}}\frac{2z_3(1-z_2)}{z_1 z_2}+\frac{4z_3}{z_1z_2}\\
  &\;\qquad\quad+(1-\eps)+\frac{4m^2}{\tilde{s}_{13}}\frac{z_3-(1-z_2)^2}{z_1z_2}
  +\frac{4m^4}{\tilde{s}_{13}\tilde{s}_{23}}\bigg\}+(1\leftrightarrow 2)\;.
  \end{split}
\end{equation}
and
\begin{equation}\label{eq:pure_split_ab_qgg}
  \begin{split}
    &\langle P^{({\rm ab},p,{\rm f})}_{q\to ggq}(p_1,p_2,p_3)\rangle=
    C_F^2(1-\eps)\bigg\{\frac{\tilde{s}_{123}^2}{2\tilde{s}_{13}\tilde{s}_{23}}\,z_3\bigg(
    \frac{(z_1+z_2)^2}{z_1 z_2}+\eps\bigg)+\bigg(\frac{s_{12}}{\tilde{s}_{13}}
    +\frac{z_1z_2}{(1-z_2)^2}\bigg)(1-\eps)\\
    &\;\qquad-\frac{\tilde{s}_{123}}{\tilde{s}_{13}}\bigg[\,
    (1-z_2)\bigg(\frac{(z_1+z_2)^2}{z_1z_2}+1\bigg)
    +(1-\eps)\bigg(\frac{z_1 z_2}{1-z_2}
    -z_3\bigg)\bigg]
    +\frac{2z_1}{z_2}\bigg(\frac{1}{1-z_2}+\frac{z_3}{1-z_1}\bigg)\\
    &\;\qquad-\frac{2m^2}{\tilde{s}_{13}}\frac{z_2(1+z_3)}{1-z_2}
    -\frac{m^2 s_{12}}{\tilde{s}_{13}\tilde{s}_{23}}(1-z_3)\bigg\}
    +\Big(1\leftrightarrow 2\Big)\;.
  \end{split}
\end{equation}

For the non-abelian quark-to-gluon splitting, we find the following decomposition~\cite{Campbell:2025lrs}
\begin{equation}\label{eq:split_qgg_nab_assembly}
  \begin{split}
    \langle P_{q\to ggq}^{\rm(nab)}(p_1,p_2,p_3)\rangle
    =&\;\frac{C_FC_A}{4}\,\tilde{s}_{123}^2\Big[\,
    \bar{\mathcal{S}}_{3;3}(p_{12};\bar{n})\Big(\mathcal{S}_{1;1}(p_2;\bar{n})
    +\mathcal{S}_{2;2}(p_1;\bar{n})
    -\mathcal{S}_{1;3}(p_2;\bar{n})-\mathcal{S}_{2;3}(p_1;\bar{n})
    \Big)\\
    &\qquad\qquad\quad\;
    +2(1-\eps)\,\mathcal{S}^{\mu\nu}_{3;3}(p_{12};\bar{n})
    D_\mu(p_1,p_2,\bar{n})D_\nu(p_1,p_2,\bar{n})\,\Big]\\
    &\;+\frac{C_A}{2}\,\tilde{s}_{123}\Big[\,
    \langle P_{q\to q}^{\rm(f)}(p_3,p_{12})\rangle
    \Big(\mathcal{S}_{1;1}(p_2;\bar{n})+\mathcal{S}_{2;2}(p_1;\bar{n})
    -\mathcal{S}_{1;3}(p_2;\bar{n})-\mathcal{S}_{2;3}(p_1;\bar{n})\Big)\\
    &\qquad\qquad\quad\;
    +2(1-\eps)\,\langle P_{q\to q}^{\rm(f)\,\mu\nu}(p_3,p_{12})\rangle
    D_\mu(p_1,p_2,\bar{n})D_\nu(p_1,p_2,\bar{n})\,\Big]\\
    &\;-\frac{C_A}{2C_F}\Big[
    P^{({\rm ab},p)}_{\tilde{q}\to gg\tilde{q}}(p_1,p_2,p_3)
    +\langle P^{({\rm ab},p,{\rm f})}_{q\to ggq}(p_1,p_2,p_3)\rangle\Big]\\
    &\;+P^{({\rm pnab},p)}_{\tilde{q}\to gg\tilde{q}}(p_1,p_2,p_3)
    +\langle P^{({\rm pnab},p,{\rm f})}_{q\to ggq}(p_1,p_2,p_3)\rangle\;.
  \end{split}
\end{equation}
where the pure scalar and fermionic components are given by
\begin{equation}\label{eq:tc_sqgg_pnab_pure}
  \begin{split}
  P^{({\rm pnab},p)}_{\tilde{q}\to gg\tilde{q}}&(p_1,p_2,p_3)=
  -\frac{C_FC_A}{4}\tilde{s}_{123}^2\Big[\,
    \mathcal{S}^{\rm(nab)}_{3;3}(p_1,p_2;\bar{n})
    +\mathcal{S}^{\rm(ab)}_{3;3}(p_1,p_2)
    +2\mathcal{S}^{\rm(ab)}_{3,3;3}(p_1,p_2;\bar{n})\\
  &\quad\;+\mathcal{S}^{\rm(ab)}_{3,3;3,3}(p_1,p_2;\bar{n})
    +\bar{\mathcal{S}}_{3;3}(p_{12};\bar{n})\Big(
    \mathcal{S}_{2;2}(p_1;\bar{n})+\mathcal{S}_{1;1}(p_2;\bar{n})
    -\mathcal{S}_{2;3}(p_1;\bar{n})-\mathcal{S}_{1;3}(p_2;\bar{n})\Big)\\
  &\quad\;+2(1-\eps)\,
    \mathcal{S}^{\mu\nu}_{3;3}(p_{12};\bar{n})\,
    D_\mu(p_1,p_2,\bar{n})D_\nu(p_1,p_2,\bar{n})\,\Big]
  +\frac{C_A}{2C_F}P^{({\rm ab},p)}_{\tilde{q}\to gg\tilde{q}}(p_1,p_2,p_3)\\
  &\qquad\qquad\;=C_FC_A\bigg\{\frac{1-\eps}{4}+
  \bigg(\frac{\tilde{s}_{123}}{\tilde{s}_{13}}\frac{z_1}{1-z_3}-1\bigg)
  \bigg(\frac{\tilde{s}_{123}}{s_{12}}\frac{1-z_1}{z_2}
  -\frac{2m^2}{s_{12}}\frac{z_1}{z_2}-1\bigg)
  -1\\
  &\;\quad+\bigg[\frac{\tilde{s}_{13}-\tilde{s}_{23}}{\tilde{s}_{13}}
  \frac{z_1-z_2}{z_2}-\frac{s_{12}}{\tilde{s}_{13}}\bigg(\frac{1-z_3}{z_2}+\frac{2z_3}{z_1}\bigg)\bigg]\frac{z_3}{(1-z_3)^2}
  +\frac{2z_3}{z_1z_2}
  -\frac{2m^2}{\tilde{s}_{13}}\frac{1}{z_2}\bigg\}+(1\leftrightarrow 2)\;.
  \end{split}
\end{equation}
and
\begin{equation}\label{eq:pure_split_pnab_qgg}
  \begin{split}
  &\langle P^{({\rm pnab},p,{\rm f})}_{q\to ggq}(p_1,p_2,p_3)\rangle
  =C_FC_A\,(1-\eps)\,\bigg\{\frac{z_2}{z_1}\bigg(\frac{1}{1-z_1}
  +\frac{z_3}{1-z_2}\bigg)+\bigg(\frac{2z_2}{z_1}+1\bigg)
  \bigg(1-\frac{3}{4}\frac{\tilde{s}_{123}}{s_{12}}(1-z_3)\bigg)\\
  &\;\qquad
  +\frac{1-\eps}{2}\bigg[\,1+\frac{s_{12}}{\tilde{s}_{13}}
    +\frac{z_1}{1-z_2}\bigg(\frac{z_2}{1-z_2}
    -\frac{\tilde{s}_{123}}{\tilde{s}_{13}}\bigg)\bigg]+\frac{\tilde{s}_{123}}{2\tilde{s}_{13}}
  \bigg(\frac{\tilde{s}_{123}}{s_{12}}-\frac{1-z_2}{z_1}\bigg)
  \bigg(\frac{(1-z_3)^2}{z_2}+\frac{z_2^2}{1-z_3}\bigg)\\
  &\;\qquad+\bigg(\frac{\tilde{s}_{123}}{s_{12}}\frac{1-z_3}{z_1}
  +\frac{\tilde{s}_{123}}{\tilde{s}_{13}}\frac{1-z_2}{z_1}
  -\frac{\tilde{s}_{123}^2}{s_{12}\tilde{s}_{13}}\bigg)
  \bigg(1-z_3-\frac{s_{12}}{\tilde{s}_{123}}\bigg)
  +\frac{m^2}{\tilde{s}_{13}}\frac{z_1-z_2}{1-z_2}-\frac{3}{4}
  \bigg\}+(1\leftrightarrow 2)\;.
  \end{split}
\end{equation}

\subsection{Gluon splitting to quarks}
\label{sec:decomposition_gluon_quark}
The abelian gluon-to-quark splitting tensor can be assembled from
lower-order components as follows
\begin{equation}\label{eq:split_gqq_ab_decomposition}
  \begin{split}
    P_{g\to gq\bar{q}}^{\mu\nu\,\rm(ab)}(p_1,p_2,p_3)=&\;
    \bigg[\frac{C_F}{2}\,s_{123}\,P_{g\to q}^{\mu\nu}(p_{12},p_3)\,
    \Big(\mathcal{S}_{2;2}(p_1;\bar{n})-\mathcal{S}_{2;3}(p_1;\bar{n})
    \Big)\\
    &\quad\;+\frac{s_{123}}{\tilde{s}_{12}}\,P_{g\to q}^{\mu\nu}(p_{12},p_3)
    \langle P^{\rm(f)}_{q\to q}(p_2,p_1)\rangle+(2\leftrightarrow 3)\,\bigg]
    +P^{\mu\nu\,({\rm ab},p)}_{g\to gq\bar{q}}(p_1,p_2,p_3)
    +\ldots\;.
  \end{split}
\end{equation}
In the massive case, the pure splitting tensor is given by
\begin{equation}\label{eq:pure_split_ab_ggq}
  \begin{split}
    &P^{\mu\nu\,({\rm ab},p)}_{g\to gq\bar{q}}(p_1,p_2,p_3)=
    C_FT_R\bigg\{-d^{\mu\nu}(p_{123},\bar{n})\bigg[
    \frac{(\tilde{s}_{123}-\tilde{s}_{23})^2}{
    2\tilde{s}_{12}\tilde{s}_{13}}(1+\eps)
    -\frac{z_1}{1-z_3}\bigg(\frac{1}{1-z_3}
    -\frac{2m^2}{\tilde{s}_{12}}\bigg)(1-\eps)\bigg]\\
    &\;\quad-d^{\mu\nu}(p_{123},\bar{n})\bigg[
    \frac{\tilde{s}_{13}}{\tilde{s}_{12}}\frac{z_2}{z_1(1-z_2)}
    +\frac{\tilde{s}_{23}}{\tilde{s}_{12}}\frac{1-2z_2}{1-z_2}
    -\frac{1}{1-z_3}\bigg(\frac{z_2}{z_1}
    -\frac{2m^2}{\tilde{s}_{12}}\bigg)+1\,\bigg]\\
    &\;\quad+\bigg(d^{\mu\nu}(p_{123},\bar{n})
    -\frac{4\tilde{p}_{3,12}^\mu\tilde{p}_{3,12}^\nu}{
    \tilde{s}_{123}}\bigg)\bigg[\,\frac{\tilde{s}_{123}}{\tilde{s}_{12}}
    \bigg(\frac{z_1-z_2}{z_1}-(1-\eps)\frac{z_1}{1-z_3}\bigg)
    +\frac{\tilde{s}_{123}}{\tilde{s}_{13}}\frac{1-z_2}{z_1}\bigg]\\
    &\;\quad+2\,\tilde{p}_{1,23}^\mu\,\tilde{p}_{1,23}^\nu\,
    \frac{\tilde{s}_{123}}{\tilde{s}_{12}\tilde{s}_{13}}\,
    \bigg(\eps-\frac{2m^2}{\tilde{s}_{123}}\,(1-\eps)\bigg)
    \bigg\}+(2\leftrightarrow 3)\;.
  \end{split}
\end{equation}
The non-abelian splitting tensor admits the following decomposition
\begin{equation}\label{eq:split_gqq_nab_assembly}
  \begin{split}
    P_{g\to gq\bar{q}}^{\mu\nu\,\rm(nab)}(p_1,p_2,p_3)
    =&\;\frac{C_A}{2}\,\frac{s_{123}^2}{s_{23}}\,\Big[\,
    \mathcal{S}^{\alpha\beta}_{1;1}(p_{23})
    P_{g\to q,\alpha\beta}(p_2,p_3)d^{\mu\nu}(p_{123},\bar{n})
    +\mathcal{S}_{23;23}(p_1;\bar{n})
    P^{\mu\nu}_{g\to q}(p_2,p_3)\,\Big]\\
    &\;+\frac{C_A}{2}\,\frac{s_{123}^2}{s_{23}}\,
    2(1-\eps)D^\mu(p_1,p_{23},\bar{n})D^\nu(p_1,p_{23},\bar{n})\,
    \langle P_{g\to q}(p_2,p_3)\rangle\\
    &\;+\frac{C_A}{4}\,s_{123}\,\Big[\,
    P_{g\to q}^{\mu\nu}(p_{12},p_3)
    +P_{g\to q}^{\mu\nu}(p_2,p_{13})\,\Big]\,
    \mathcal{S}_{2;3}(p_1;\bar{n})\\
    &\;-\frac{C_A}{2C_F}\,
    P^{\mu\nu\,({\rm ab},p)}_{g\to gq\bar{q}}(p_1,p_2,p_3)
  +P^{\mu\nu\,({\rm pnab},p)}_{g\to gq\bar{q}}(p_1,p_2,p_3)+\ldots\;.
  \end{split}
\end{equation}
where the pure splitting tensor is given by
\begin{equation}\label{eq:pure_split_gqqb_pnab_2}
  \begin{split}
    &P^{\mu\nu\,({\rm pnab},p)}_{g\to gq\bar{q}}(p_1,p_2,p_3)=
    \frac{C_AT_R}{2}\bigg\{-d^{\mu\nu}(p_{123},\bar{n})\,
    (1-\eps)\bigg(\frac{s_{123}}{\tilde{s}_{12}}\frac{z_1}{1-z_3}
    -\frac{\tilde{s}_{13}}{\tilde{s}_{12}}-\frac{z_1}{(1-z_3)^2}\bigg)\\
    &\;\quad-d^{\mu\nu}(p_{123},\bar{n})\bigg[
    \frac{1-2z_1}{z_1}\bigg(\frac{s_{123}}{\tilde{s}_{12}}\frac{1-z_3}{1-z_1}
    +\frac{s_{123}}{s_{23}}-\frac{s_{123}^2}{\tilde{s}_{12}s_{23}}\frac{z_2}{1-z_1}\bigg)
    -\frac{2}{z_1}+1-\frac{1}{1-z_3}\bigg(\frac{2z_2}{z_1}-\frac{2m^2}{\tilde{s}_{12}}\bigg)\bigg]\\
    &\;\quad+\frac{4\tilde{p}_{1,23}^\mu\tilde{p}_{1,23}^\nu}{s_{23}}
    \bigg[\,\frac{2z_2z_3}{(1-z_1)^2}-(1-\eps)
    +\frac{4m^2}{\tilde{s}_{13}}\frac{z_2}{1-z_1}\bigg]
    +\frac{2(\tilde{p}_{2,13}^\mu\tilde{p}_{3,12}^\nu
    +\tilde{p}_{3,12}^\mu \tilde{p}_{2,13}^\nu)}{s_{23}}
    \frac{s_{123}}{\tilde{s}_{12}}\bigg[\frac{2z_2}{z_1}\frac{z_3-z_1}{1-z_1}+1-\eps\bigg]\\
    &\;\quad-\frac{4\tilde{p}_{3,12}^\mu\tilde{p}_{3,12}^\nu}{s_{23}}\bigg[\,\frac{2z_2}{z_1}
    \bigg(\frac{s_{123}}{\tilde{s}_{12}}\frac{z_2}{1-z_1}-\frac{s_{23}}{\tilde{s}_{12}}\bigg)
    -(1-\eps)\bigg(\frac{s_{123}}{\tilde{s}_{13}}
    +\frac{s_{23}}{\tilde{s}_{12}}\frac{z_1}{1-z_3}\bigg)\bigg]\\
    &\;\quad-\frac{8}{z_1}\frac{\tilde{p}_{2,3}^\mu\tilde{p}_{2,3}^\nu}{s_{23}}
    +\frac{8m^2}{\tilde{s}_{12}}\frac{\tilde{p}_{1,23}^\mu\tilde{p}_{2,3}^\nu
    +\tilde{p}_{1,23}^\nu\tilde{p}_{2,3}^\mu}{s_{23}}\bigg\}+(2\leftrightarrow 3)\;.
  \end{split}
\end{equation}
The ellipses in Eqs.~\eqref{eq:split_gqq_ab_decomposition}
and~\eqref{eq:split_gqq_nab_assembly} stand for terms proportional
to $\bar{n}^\mu$ or $\bar{n}^\nu$, which vanish when contracted with
a gluon polarization tensor.

\subsection{Kinematical limits of pure splitting remainders}
\label{sec:kinematical_limits}
In Ref.~\cite{Campbell:2025lrs}, it was found that the massless pure splitting
remainders corresponding to $\langle P_{q\to q\bar{q}q}^{(p)}\rangle$,
$\langle P_{q\to ggq}^{({\rm ab},p,f)}\rangle$,
$\langle P_{q\to ggq}^{({\rm pnab},p,f)}\rangle$,
$P_{g\to gq\bar{q}}^{\mu\nu\,({\rm ab},p)}$, and
$P_{g\to gq\bar{q}}^{\mu\nu\,({\rm pnab},p)}$ in
Eqs.~\eqref{eq:pure_split_qbqq}, 
\eqref{eq:pure_split_ab_qgg},
\eqref{eq:pure_split_pnab_qgg},
\eqref{eq:pure_split_ab_ggq},
and~\eqref{eq:pure_split_gqqb_pnab_2}
have at most sub-leading singularities in the soft-gluon limits (if applicable)
and in the collinear limits. This statement still holds for the massive pure
splitting remainders, if the collinear limit is replaced by the quasi-collinear one.
In particular, Tab.~I of Ref.~\cite{Campbell:2025lrs} remains unchanged 
if the collinear limit is replaced by the quasi-collinear limit,
and the massive extensions of the pure splitting functions are used.
In complete analogy, the functions
$P_{\tilde{q}\to gg\tilde{q}}^{({\rm ab},p)}$ and
$P_{\tilde{q}\to gg\tilde{q}}^{({\rm pnab},p)}$ in
Eqs.~\eqref{eq:pure_split_ab_sqgg}
and~\eqref{eq:tc_sqgg_pnab_pure} have leading soft singularities, which arise
from the interference contributions of the leading diagrams in the axial gauge.
The scalar splitting functions should therefore be expressed in terms of the
scalar radiator functions by making use of Eqs.~\eqref{eq:scalar_tc_correspondence}.
In summary, we find that the methodology for the computation of splitting functions
first introduced in Ref.~\cite{Campbell:2025lrs} applies to both massive and
massless QCD, because it is based on a physical, rather than a purely mathematical
understanding of the origin of infrared singularities.

\section{Conclusions}
\label{sec:outlook}
Using a recently introduced method for the decomposition of splitting functions
into scalar dipole radiators, lower-order expressions, and pure splitting remainders,
we derived a representation of the massive one-to-three-parton splittings that makes
their factorization manifest. The massive two-gluon radiators used in this context
were computed in a process-independent form for the first time. The pure splitting
remainders exhibit at most sub-leading singularities in the (double-)soft gluon limits
and in the quasi-collinear limits. Kinematical limits are, however, not used in any
of our calculations, which implies that the results remain valid across the complete
physical phase space. 

As a byproduct of our analysis, we derived the massive QCD one-to-three parton
splitting functions in a significantly more compact form than existing literature.
This will reduce evaluation time and increase numerical stability in practical
applications. We have also obtained the first independent cross-check of the massive,
triple-collinear splitting functions that do not exist in the SCET-based approach
of~\cite{Craft:2023aew}, in particular Eqs.~(5.9) and~(5.10) of~\cite{Dhani:2023uxu}.

We expect the novel techniques introduced in this manuscript to provide
a path towards higher-order computations which do not require sectorization
or other forms of overlap removal for different singular phase-space configurations.
At the same time, the splitting functions derived here will be helpful
for higher logarithmic resummation of soft gluon effects on QCD observables.
These methods will help drive improvements in the modeling of heavy quark
jet formation, leading to better precision in analyses of top quark and
electroweak boson production at the LHC.

\section*{Acknowledgments}
\noindent
This manuscript has been authored by Fermi Forward Discovery Group, LLC
under Contract No. 89243024CSC000002 with the U.S.\ Department of Energy,
Office of Science, Office of High Energy Physics.
The work of S.H. was supported by the U.S. Department of Energy,
Office of Science, Office of Advanced Scientific Computing Research,
Scientific Discovery through Advanced Computing (SciDAC-5) program,
grant “NeuCol”.
M.L.B., J.R. and G.W. gratefully acknowledge the support of the
Brown University Department of Physics and Graduate School.
Finally, M.L.B. and G.W. would like to thank David~Axen for his prescient gift
of a copy of R.~D.~Field's \emph{Applications of Perturbative QCD}~\cite{Field:1989uq}
that was of instrumental value to this project.

\bibliography{main}
\end{document}